\documentclass[fleqn,opre,nonblindrev]{informs_modified}

\DoubleSpacedXI

\usepackage{endnotes}
\let\footnote=\endnote

\usepackage[small, margin=1cm]{caption}

\usepackage{appendix}
\usepackage{color}
\definecolor{strcolor}{rgb}{0.6, 0.2, 0.6}
\definecolor{commentcolor}{rgb}{0.3125, 0.5, 0.3125}
\definecolor{keycol}{rgb}{0, 0, 1}

\usepackage{natbib}
\bibpunct[, ]{(}{)}{,}{a}{}{,}%
\def\bibfont{\fontsize{8}{9.5}\selectfont}%
\def\newblock{\ }%
\def\BIBand{and}%

\TheoremsNumberedThrough     
\ECRepeatTheorems

\EquationsNumberedBySection 

\MANUSCRIPTNO{ }

\usepackage{amssymb}
\usepackage{amsmath}
\usepackage{indentfirst}
\usepackage{graphicx}
\usepackage[colorlinks=true, allcolors=black]{hyperref}
\usepackage[utf8]{inputenc} 
\usepackage[T1]{fontenc}    
\usepackage{url}            
\usepackage{booktabs}       
\usepackage{amsfonts}       
\usepackage{nicefrac}       
\usepackage{microtype}      
\usepackage{fancyhdr}       
\graphicspath{{media/}}

\usepackage{subcaption}

\newcommand{\ud}{\mathrm{d}}

\newcommand{\E}{\mathbb{E}}

\newcommand{\F}{\mathcal{F}}

\newcommand{\R}{\mathbb{R}}

\newcommand{\one}{\mathbf{1}}

\begin{document}


\RUNAUTHOR{Li, Liang and Pang}

\RUNTITLE{Continuous-Time Monotone Mean-Variance Portfolio Selection in Jump-Diffusion Model}

\TITLE{Continuous-Time Monotone Mean-Variance Portfolio Selection in Jump-Diffusion Model}

\ARTICLEAUTHORS{%
\AUTHOR{Yuchen Li, Zongxia Liang}
\AFF{Department of Mathematical Sciences, Tsinghua University, Beijing 100084, China, \EMAIL{li-yc21@mails.tsinghua.edu.cn}, \EMAIL{liangzongxia@tsinghua.edu.cn}}
\AUTHOR{Shunzhi Pang}
\AFF{School of Economics and Management, Tsinghua University, Beijing 100084, China, \EMAIL{psz22@mails.tsinghua.edu.cn}}
} 

\ABSTRACT{%
We study continuous-time portfolio selection under monotone mean-variance (MMV) preferences in a jump-diffusion model, presenting an explicit solution different from that under classical mean-variance (MV) preferences in dynamic settings for the first time. We prove that the potential measures calculating MMV preferences can be restricted to non-negative Doléans-Dade exponentials. We find that MMV can resolve the non-monotonicity and free cash flow stream problems of MV when the jump size can be larger than the inverse of the market price of risk. Such result is completely comparable to the earliest result by Dybvig and Ingersoll. Economically, we show that the essence of MMV lies in the pricing operator always remaining non-negative, with a value of zero assigned when the jump exceeds a certain threshold, avoiding the issue of non-monotonicity. As a result, MMV investors behave markedly different from MV investors. Furthermore, we validate the two-fund separation and establish the monotone capital asset pricing model (monotone CAPM) for MMV investors. We also study MMV in a constrained trading model and provide three specific numerical examples to show MMV's efficiency. Our finding can serve as a crucial theoretical foundation for future empirical tests of MMV and monotone CAPM's effectiveness.
}%






\maketitle

%


\section{Introduction}{\label{Section Introduction}}
\noindent
Mean-variance (MV) preference is one of the most widely used portfolio selection criteria in both theoretical finance research and practical investment \citep{markowitz1952portfolio, zhou2000continuous, basak2010dynamic, dai2021dynamic}. However, due to the well-known non-monotonicity of MV preference \citep{dybvig1982mean, jarrow1997mean}, the monotone mean-variance (MMV) preference proposed by \citet{maccheroni2009portfolio} is becoming widely concerned. Theoretically, MMV represents a specific type of the variational preference proposed by \citet{maccheroni2006ambiguity, maccheroni2006dynamic}, serving as an optimal monotone approximation of classical MV and enhancing the economic significance of MV portfolio selection. However, due to the lack of explicit solutions for MMV portfolio selection distinct from those under MV, its practical application has been limited. Apart from the initial single-period example provided by \citet{maccheroni2009portfolio}, there are currently no explicit solutions under dynamic settings showing the difference of MV and MMV. In this paper, we address this theoretical gap by providing a solution for optimal MMV portfolio selection within a continuous-time jump-diffusion financial market. The solution is presented in an explicit form and can be different from that under MV preferences under a certain condition. 

In the literature, several works have studied MMV portfolio selection problems across various market models, and all have found solutions consistent with those under classical MV preferences \citep{trybula2019continuous, li2021optimal, shen2022cone, hu2023constrained}. Proving that both wealth processes of MV and MMV investors under the optimal strategies would not leave the domain of monotonicity of MV preferences, \citet{strub2020note} and \citet{du2023monotone} generalize the conclusion that the consistency of MV and MMV holds whenever asset prices are continuous, regardless of whether there are trading constraints in the market. Thus, discontinuity emerges as a necessary condition for MV and MMV to be different. 

\citet{li2022comparison} study the MMV portfolio selection problem in a discontinuous market for the first time. In an extended model of \citet{trybula2019continuous}, they compare the solutions to MV and MMV problems and propose a sufficient and necessary condition for them to be consistent. However, due to the complexity of the stochastic factor market model, they do not fully resolve the MMV problem when it differs from the MV problem. Based on the classical Cramér-Lundberg insurance model, \citet{li2023optimal} find the consistency of MV and MMV still holds even when there are jumps in the risk model. This finding indicates that there may be more fundamental reasons beyond discontinuity causing the inconsistency between MV and MMV. 

Indeed, the findings of \citet{vcerny2020semimartingale} have preliminarily shed light on this potential issue. In a general semimartingale model, \citet{vcerny2020semimartingale} represents solutions to MV and MMV problems symbolically and finds that their consistency is contingent upon the variance-optimal martingale measure being unsigned. When asset prices are continuous, the variance-optimal martingale measure remains non-negative. In the presence of jumps in asset prices, the density of the variance-optimal measure may be signed \citep{schweizer1994approximating, schweizer1996approximation}. However, \citet{vcerny2020semimartingale} does not give an explicit solution to the MMV investment problem, which is one of our main considerations. Based on our explicit solution, the phenomenon of the signed measure can be observed clearly. 

In this paper, we set a classical jump-diffusion financial market model and solve the optimal MMV portfolio selection. Unlike the conventional approach that relies on the consistency of MV and MMV, which does not hold in our jump-diffusion model, we first introduce a theorem demonstrating that the potential measures calculating MMV can be restricted to the set of all non-negative Doléans-Dade exponentials. This enables an easier characterization of MMV in dynamic settings, helping us solve the portfolio selection problem directly. Next, the problem can be regarded as a stochastic differential game between the investor and the market player. We solve it with the HJBI equation and prove the validity of a verification theorem. Finally, we get the solution in an explicit form and compare it with that of MV problem. Similar to \citet{li2022comparison}, we prove that the two optimal strategies and value functions coincide if and only if the financial market admits that $\Delta{L} \leq q^{mv}$ almost surely (a.s.), where $L$ represents the L\'{e}vy process describing the jump in the market, $\Delta{L}$ denotes the corresponding jump size, and $q^{mv}$ is the inverse of the market price of risk. Hence, when the potential jump size surpasses a certain threshold, the behavior of MMV investors would diverge from that of MV investors. 

Our finding reveals an intriguing divergence between MMV and MV investors. Specifically, we observe that MMV investors tend to maintain a zero position in the risky asset when the jump size exceeds a certain threshold. In contrast, MV investors have a pre-determined expected wealth value and always follow a strategy linear in current wealth to reach this target. In the event of a significant upward jump causing wealth to surpass this preset target, MV investors persist in pursuing this target, even if it means assuming a short position and incurring negative profits. This seemingly irrational strategy stems from MV preferences treating large upward profits as risks due to the variance part, thus leading to the non-monotonicity. 

We also investigate the relationship between the wealth process and the domain of monotonicity of MV. When the market satisfies $\Delta{L} \leq q^{mv}$ a.s., both the terminal wealth of MV and MMV investors remain within the monotone domain. Consequently, MV and MMV keep consistent. However, when $\Delta{L} \leq q^{mv}$ a.s. does not hold, there exists a positive probability that the terminal wealth of both investors fall outside the monotone domain. In such instances, MV emerges the free cash flow streams (FCFS) problem \citep{cui2012better}, while MMV fixes this problem due to its monotonicity. All these evidences suggest that MMV presents a more rational approach compared to MV. 

As one of the central focuses of this paper, we reveal the economic essence of MMV by studying the pricing operator taken by investors, which also serves as the fundamental reason for the inconsistency of MV and MMV. We find that the MV investor optimizes his portfolio under a pricing operator which can be negative when $\Delta{L} > q^{mv}$. However, the pricing operator of the MMV investor is always non-negative. Specifically, it is assigned a value of zero when the jump exceeds a certain threshold, effectively avoiding the problem of non-monotonicity. Besides, when $\Delta{L} \leq q^{mv}$ a.s. holds, the variance-optimal martingale measure in this market remains unsigned, verifying the conclusion of \citet{vcerny2020semimartingale}. Furthermore, our findings in the continuous-time scenario are strikingly consistent with \citet{dybvig1982mean}, where the non-monotonicity of MV was initially proposed in a discrete-time model. Our derived consistency condition $\Delta{L} \leq q^{mv}$ a.s. is consistent with $\widetilde{X}_m \leq \bar{X}_m + \frac{1}{\lambda_m}$ a.s. proposed by \citet{dybvig1982mean}, where $\widetilde{X}_m$ represents the potential market return per unit of capital and $\lambda_m$ denotes the market price of risk. 

To further refine the continuous-time MMV portfolio selection theory, we extend our results from the single-asset market to the multi-asset market. Employing a similar methodology, we solve the MMV problem explicitly and provide a sufficient and necessary condition for MV and MMV to be consistent. Our analysis reveals the existence of a market portfolio, suggesting that investors with varying levels of uncertainty aversion should hold a combination of the risk-free asset and the market portfolio. Thus, the famous two-fund separation theorem remains valid \citep{sharpe1964capital, lintner1965valuation, mossin1966equilibrium}. 

Based on the market portfolio, we establish the monotone capital asset pricing model (monotone CAPM). The risk premium of each asset still has a linear relationship with the risk premium of the market portfolio, indicating that the idiosyncratic risk can be fully diversified and only the systematic risk should be priced. However, the beta is modified due to the different pricing operator and the different market portfolio under MMV. Notably, the monotone CAPM ensures the arbitrage-free condition, distinguishing it from the classical CAPM. These findings align with those of the single-period model \citep{maccheroni2009portfolio}, providing a comprehensive understanding across different temporal frameworks. Moreover, parameters in dynamic settings can be estimated more precisely, which in turn facilitates future empirical tests of MMV and Monotone CAPM's effectiveness. 

Furthermore, inspired by \citet{du2023monotone}, we study MMV portfolio selection under a no-short constraint and compare it with the original model. While the MV optimal strategy changes when a jump larger than $q^{mv}$ occurs, the MMV optimal strategy remains unchanged compared to the market without the no-short constraint. Once again, we demonstrate that the consistency between MV and MMV is contingent upon $\Delta{L} \leq q^{mv}$ a.s., even in a market with a no-short trading constraint. 

\begin{figure}[htbp]
    \FIGURE
    {\includegraphics[width=1\linewidth]{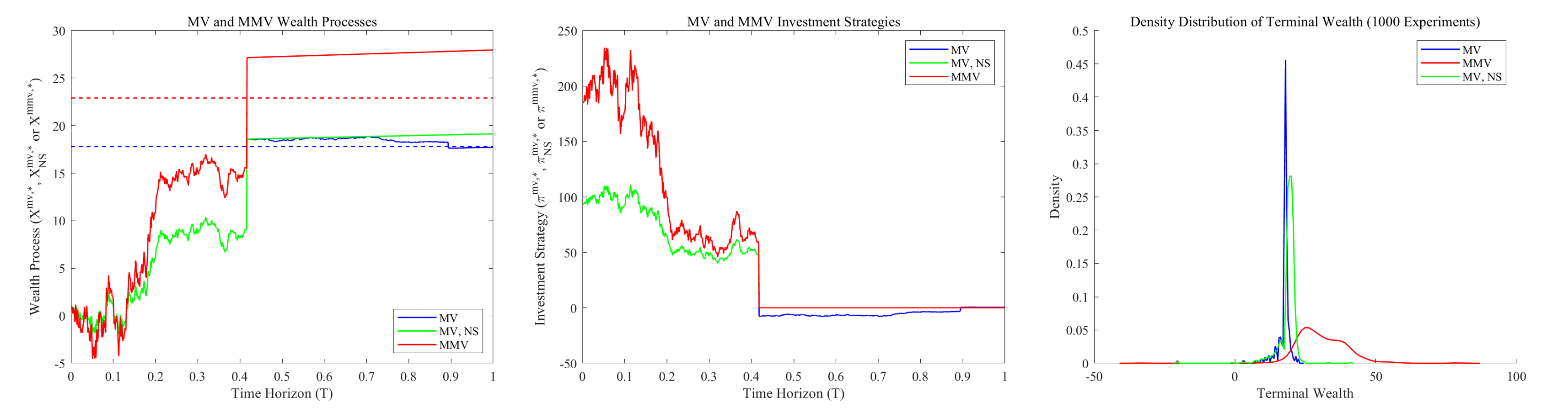}}
    {Difference Between MV and MMV \label{Intro Constant Distributed}}
    {The first sub-figure shows the wealth process under the MV (MMV) optimal strategy, the second sub-figure shows the corresponding optimal strategy, and the third sub-figure shows the estimated density distribution (by 1000 experiments) of investor's terminal wealth. The solid blue curve represents the MV strategy, the green one signifies MV with no-short constraint, and the red curve corresponds to MMV. The blue (red) dashed curve represents the pre-determined target of wealth that drives the investor to short sell (hold zero position) of the risky asset under MV (MMV).} 
\end{figure}

To provide a more direct demonstration of MMV's efficiency, we present three specific examples comparing MV and MMV numerically. These examples involve constantly, uniformly, and exponentially distributed jumps. Through numerical simulations, we observe distinct patterns in wealth processes and investment strategies under MV, MV with no-short constraint, and MMV. Also, the distribution of end-of-period wealth under MMV appears to have a higher mean and a thicker tail due to its monotonicity property. Figure \ref{Intro Constant Distributed} shows an example when MMV differs from MV, in which we can observe different behaviors between MV and MMV investors. These observations further validate our theoretical findings. 

To summarize, we contribute to the literature in at least four aspects. First, we provide an explicit solution to continuous-time MMV portfolio selection problem that is different from MV for the first time. Such result not only enriches current theoretical literature \citep{trybula2019continuous, li2022comparison}, but also provides a basis for future empirical tests of MMV's effectiveness. 

Second, based on our solution, we interpret its economic implications and shed light on the fundamental reason behind MMV's resolution of the non-monotonicity. The MMV investor adopts a pricing operator that always remains non-negative, thus avoiding the problem of non-monotonicity. Also, it tends to maintain a zero position of the risky asset indefinitely after the jump size exceeds a certain threshold, which is a significant departure from the behavior of the MV investor. This result not only aligns seamlessly with \citet{dybvig1982mean}'s seminal work, but also complements discussions on the existence of FCFS in financial markets \citep{cui2012better, bauerle2015complete, strub2020note, vcerny2020semimartingale}. 

Third, we discover the two-fund separation theorem for MMV investors within a continuous-time multi-asset model and establish the monotone CAPM with an explicit expression.This finding mirrors outcomes from the single-period case \citep{maccheroni2009portfolio}, enabling more precise estimation of relevant market parameters. 

Last, we introduce a new theorem demonstrating that the target measures of MMV can be restricted on the set of all the non-negative Doléans-Dade exponentials. This method enables MMV to be characterized more easily in dynamic settings, opening avenues for tackling more MMV problems in a similar way. 

The remainder of the paper is structured as follows. Section \ref{Section Model} outlines the market model setup. Section \ref{Section MV Portfolio Selection} solves the classical MV problem. Section \ref{Section MMV Portfolio Selection} proposes a modification of MMV problem and formulates the HJBI equation to solve it. In Section \ref{Section Consistency}, we compare MV and MMV and reveal the economic essence of MMV. Section \ref{Section Multi-Asset} extends MMV portfolio selection to a multi-asset model, verifies the two-fund separation theorem, and establishes the monotone CAPM. Section \ref{Section Constraint} resolves MMV problem under the trading constraint. Section \ref{Section Numerical} provides three specific examples to compare MV and MMV numerically. Finally, Section \ref{Section Conclusion} concludes the paper. 

\section{General Model and Assumption}{\label{Section Model}}
\noindent
Consider a complete probability space $(\Omega,\mathcal{F},\{\mathcal{F}_t\}_{t\geq0},\mathbb{P})$ satisfying the usual conditions and filtration $\{\mathcal{F}_t\}_{t\geq0}$ is generated by $B$ and $N$. Here, $B \triangleq \Big\{B(t);t \geq 0\Big\}$ is a one-dimensional standard Brownian motion, and ${N}$ is a Poisson random measure on $\big([0,T] \times \mathbb{R}, \mathcal{B}\left(([0,T] \times \mathbb{R} \right)\big)$ with respect to a L\'{e}vy process $L$. All stochastic models describing the financial market are well-defined within this space.

Suppose there exists a jump-diffusion financial market where a risk-free asset (bond) and a risky asset (stock) are traded. The price processes of the two assets, $S_0 \triangleq \Big\{ S_0(t);t \geq 0\Big\}$ and $S_1 \triangleq \Big\{ S_1(t);t \geq 0\Big\}$, are given by the following stochastic differential equations (SDEs):
\begin{equation}
    \left\{
    \begin{aligned}
    & {\mathrm{d}S_0(t)} ={S_0(t)} r \mathrm{d} t, \\
    & \mathrm{d}S_1(t) ={S_1(t^-)}\Bigg \{ \mu \mathrm{d}t + \sigma \mathrm{d} B(t) + \mathrm{d} \sum_{i = 1}^{n(t)}Q_i\Bigg  \},
    \end{aligned}
    \right. \label{Jump-Diffusion Financial Market}
\end{equation}
where the risk-free interest rate $r$, the drift term $\mu$ and the volatility term $\sigma$ are all assumed to be constants for simplicity. 
The stock has a positive instantaneous expected rate of return, i.e., $\mu-r+\lambda\xi_1 > 0$. Additionally, the market is considered frictionless, with no transaction costs, no taxes, infinitely divisible shares, and no trading constraints. In Section \ref{Section Constraint}, we will explore relaxing the assumption of no trading constraints.

Given the potential for significant fluctuations in stock prices over short time intervals, we employ a compound Poisson process $L \triangleq \Big\{ L(t) = \sum\limits_{i = 1}\limits^{n(t)}Q_i; t\geq 0 \Big\}$, independent of $B$, to describe the discontinuous change in the stock's return. $\Delta{L}(t) = L(t) - L(t^-)$ is the size of the jump at time $t$. $n(t)$ is the number of jumps occurring during the time $[0,t]$, following a homogeneous Poisson process with intensity $\lambda$. $Q_i, i = 1,2,\cdots,$ are sizes of jumps and are assumed to be independently and identically distributed on $[-1, \infty)$ to guarantee the non-negativity of the stock price. Suppose that  $Q_i$ and $ Q$ have a  common distribution with finite moments $\xi_1 = \mathbb{E}^\mathbb{P}[Q]$, $\xi^2_2 = \mathbb{E}^\mathbb{P} [Q^2]$. Also, we make a technical assumption that $\mathbb{E}^\mathbb{P} [Q^8]<\infty$, which would be used in the proof of verification theorem. Most of the known distributions satisfy this assumption.
Also, let $\widetilde{N}(\mathrm{d}s, \mathrm{d}q) = N(\mathrm{d}s,\mathrm{d}q) - \nu(\mathrm{d}q)\mathrm{d}s$ be the compensated random measure and $\nu(\mathrm{d}q)$ be the L\'{e}vy measure, i.e., $L(t) = \int_0^t \int_{-1}^{\infty} q N(\mathrm{d}s,\mathrm{d}q)$ and $\mathbb{E}^\mathbb{P}[L(t)] = t\int_{-1}^{\infty}q\nu(\mathrm{d}q) = t\lambda\xi_1$. 

Suppose an investor allocates $\pi(t)$ amount into the risky asset at time $t$. $X^\pi \triangleq \Big\{X^\pi(t);t\geq 0 \Big\}$ is the investor's wealth process under the investment strategy $\pi \triangleq \Big\{ \pi(t);t\geq0\Big\}$. And the rest of the wealth $X^\pi(t) - \pi(t)$ is invested into the risk-free asset. As a result, the wealth process follows 
\begin{align}
    \mathrm{d} X^{\pi} (t) & = \pi(t) \frac{\mathrm{d}S_1 (t)}{S_1 (t^-)} + (X^{\pi}(t) - \pi(t) ) \frac{\mathrm{d}S_0 (t)}{S_0 (t)} \mathrm{d}t \notag \\ 
     & = \Big[X^{\pi} (t)r + \pi(t)(\mu-r + \lambda \xi_1) \Big]\mathrm{d}t + \pi(t)\sigma\mathrm{d}B(t) + \pi(t^-) \int_{-1}^{\infty} q \widetilde{N}(\mathrm{d}t, \mathrm{d}q). \label{Jump-Diffusion Wealth Process}
\end{align} 
This SDE represents the dynamics of the wealth process, incorporating both the risky and risk-free assets along with a jump-diffusion component.

\section{MV Portfolio Selection}{\label{Section MV Portfolio Selection}}
\noindent
Before solving the MMV problem, we derive the classical MV portfolio selection as a preliminary. Following the pre-commitment method proposed by \citet{zhou2000continuous}, we transform the problem into a stochastic LQ control problem and take the dynamic programming method to solve it. Here, we present the anticipated outcomes upfront. 

Assume that the investor has the classical MV preference:
\begin{equation}
    U_{\gamma}\left(X\right) = \mathbb{E}^\mathbb{P}\left[X\right]-\frac{\gamma}{2}\mathrm{Var}^{\mathbb{P}}\left[X\right], \quad X \in \mathcal{L}^2(\Omega,\mathcal{F},\mathbb{P}), \label{MV Definition}
\end{equation}
where $\gamma > 0$ is the uncertainty aversion coefficient. For simplicity, we denote $\mathcal{L}^2(\Omega,\mathcal{F},\mathbb{P}) = \mathcal{L}^2(\mathbb{P})$. $U_\gamma$ is not a monotone preference, i.e., there exist $X, Y \in \mathcal{L}^2(\mathbb{P})$, such that $X(\omega) \leq Y(\omega)$ for any $\omega \in \Omega$ but $U_{\gamma}\left(X\right) > U_{\gamma}\left(Y\right)$, violating the basic tenet of economic rationality. Despite this, MV has been widely studied and used in practice due to its concise form. For an MV investor with initial wealth $x$ at time $t$, its optimization problem is formulated as:

\begin{problem} \label{MV Problem}
    Under MV preferences \eqref{MV Definition}, the investor's optimization goal is to solve 
    \begin{equation}
        \sup_{\pi \in \mathcal{A}^{mv}_t}J(t,x,\pi) \triangleq  \sup_{\pi \in \mathcal{A}^{mv}_t} \left\{\mathbb{E}_{t,x}\left[X^{\pi}(T)\right] - \frac{\gamma}{2}\mathrm{Var}_{t,x}\left[X^\pi(T)\right]\right\}, \label{MV Objective Function}
    \end{equation}
    where a trading strategy $\pi \triangleq \Big\{\pi(s); s \in [t,T] \Big\}$ is admissible and denoted as $\pi \in \mathcal{A}^{mv}_t$ if it satisfies: (1) $\pi$ is predictable with respect to the filtration $\left\{  \mathcal{F}_{s}\right\}_{s\in[t,T]}$. (2) For any $(t,x) \in [0,T] \times \mathbb{R}$, $\mathbb{E}_{t,x}\left[ \sup_{t \leq s \leq T} |X^\pi(s)|^2 \right] < \infty$. (3) SDE \eqref{Jump-Diffusion Wealth Process} has a unique solution. 
\end{problem} 

Problem \ref{MV Problem} can be converted to an LQ control problem under the pre-commitment and the following theorem provides the optimal control and value function. 

\begin{theorem} \label{Theorem MV Investor}
    For an investor with initial wealth $x$ at time $t$, the optimal investment strategy under MV preferences is
    \begin{equation}
        \pi^{mv,\ast}(s) = \frac{1}{q^{mv}}\left(e^{(s-t)r}x + \frac{e^{(T-t)C^{mv}-(T-s)r}}{\gamma} -X^{mv,\ast}(s^-)\right), \quad t \leq s \leq T, \label{MV Strategy} 
    \end{equation}
    where $q^{mv} = \frac{\sigma^2 + \lambda \xi_2^2}{\mu-r+\lambda \xi_1}$, $C^{mv} = \frac{(\mu-r+\lambda \xi_1)^2}{\sigma^2 + \lambda \xi_2^2}$, and $X^{mv,\ast}$ denotes the wealth process under the strategy $\pi^{mv,\ast}$. The maximum utility gained by the MV investor is
    \begin{equation}
        U_\gamma\left(X^{mv,\ast}(T)\right) = e^{(T-t)r}x + \frac{e^{(T-t)C^{mv}}-1}{2\gamma}. 
    \end{equation}
\end{theorem} 

We opt to omit the proof, given the extensive study of analogous problems in existing literature. For instance, \citet{liang2016optimal} solve the optimal MV reinsurance-investment problem in the jump-diffusion setting. Leaving out the insurance market in their model leads to the result we need. Moreover, we give the following propositions derived from Theorem \ref{Theorem MV Investor}, aiding in a deeper understanding of the MV investor's behavior.

\begin{proposition} \label{Proposition MV Stopping Time}
    Define a stopping time $T_{q^{mv}} \triangleq \inf \Big\{s > t: \Delta{L}(s) \geq q^{mv} \Big\}$, and $\inf\{ \varnothing\}=\infty$. When $t \leq s < T_{q^{mv}}$, we have $\pi^{mv,\ast}(s) > 0$, which means that the investor holds a positive amount of the risky asset. When $ s = T_{q^{mv}}$, we have $\pi^{mv,\ast}(s) \leq 0$, which means that the investor changes to short sell or keep zero position of the risky asset. 
\end{proposition} 

\textbf{Proof.} 
    Under the optimal strategy $\pi^{mv,\ast}$, the wealth process follows
\begin{equation}
    \mathrm{d} X^{mv,\ast}(s) = X^{mv,\ast}(s)r\mathrm{d}s + \left(e^{(s-t)r}x + \frac{e^{(T-t)C^{mv}-(T-s)r}}{\gamma} - X^{mv,\ast}(s)\right) \Big(C^{mv} \mathrm{d}s + \mathrm{d}D(s) \Big), \notag 
\end{equation}
where $\mathrm{d}D(s) = \frac{\sigma}{q^{mv}} \mathrm{d}B(s) + \int_{-1}^{\infty} \frac{q}{q^{mv}} \widetilde{N}(\mathrm{d}s, \mathrm{d}q)$. By simple transformation, we have
\begin{align}
    & \mathrm{d} \left(e^{(T-t)r}x + \frac{e^{(T-t)C^{mv}}}{\gamma} - e^{(T-s)r}X^{mv,\ast}(s) \right) \notag \\
    = & \left(e^{(T-t)r}x + \frac{e^{(T-t)C^{mv}}}{\gamma} - e^{(T-s)r}X^{mv,\ast}(s) \right) \Big(-C^{mv} \mathrm{d}s - \mathrm{d}D(s) \Big). \notag
\end{align} 
When $t \leq s < T_{\bar{q}}$, $e^{(T-t)r}x + \frac{e^{(T-t)C^{mv}}}{\gamma} - e^{(T-s)r}X^{mv,\ast}(s) > 0$ and therefore $\pi^{mv,\ast}(s) > 0$. When $s = T_{q^{mv}}$, $e^{(T-t)r}x + \frac{e^{(T-t)C^{mv}}}{\gamma} - e^{(T-s)r}X^{mv,\ast}(s) \leq 0$ and $\pi^{mv,\ast}(s) \leq 0$. \hfill $\square$
\vskip 5pt

\begin{proposition} \label{Proposition MV Expression}
    When $t \leq s \leq T$, $X^{mv, \ast}(s)$ can be expressed as
    \begin{align}
        X^{mv,\ast}(s) & = e^{(s-t)r}x + \frac{e^{(T-t)C^{mv}-(T-s)r}}{\gamma} \notag \\
        & - \frac{1}{\gamma} e^{(T-s)(C^{mv}-r)} \mathcal{E} \Bigg( -\int_t^. \frac{\sigma}{q^{mv}} \mathrm{d}B(\tau) - \int_t^.\int_{-1}^{\infty} \frac{q}{q^{mv}} \widetilde{N}(\mathrm{d}\tau,\mathrm{d}q) \Bigg)_s, \notag
    \end{align}
    where $\mathcal{E}(\cdot) \triangleq \Big\{\mathcal{E}(\cdot)_s: s \geq t \Big\}$ is the Doléans-Dade exponential process of a semi-martingale with an initial value of 1. 
\end{proposition}

Based on the proof of Proposition \ref{Proposition MV Stopping Time}, the expression in Proposition \ref{Proposition MV Expression} can be naturally derived. See Section 8 of \citet{Protter2005} for more details about the Doléans-Dade exponentials. Also refer to \citet{li2022comparison} for the application of the Doléans-Dade exponentials in MV and MMV problems. 


\section{MMV Portfolio Selection}{\label{Section MMV Portfolio Selection}}
\noindent
In this section, we derive the optimal MMV portfolio selection problem, which is the main focus of this paper. Our method is similar to \citet{mataramvura2008risk} and \citet{trybula2019continuous} but with some notable differences. 
First, to characterize MMV preferences in the dynamic setting explicitly, we propose a theorem to show that the target ambiguity measures can be restricted on the set of all the non-negative Doléans-Dade exponentials. Then, we regard the MMV investment problem as a max-min problem, and give the corresponding admissible set, solution to the HJBI equation and verification theorem. 

\subsection{Objective Function}
\noindent
Now we assume that the investor has the MMV preference given by
\begin{equation}
    V_{\gamma}(X)=\inf_{\mathbb{Q} \in \Delta^2(\mathbb{P})} \left\{ \mathbb{E^{Q}}[X]+\frac{1}{2\gamma}\mathbb{C}(\mathbb{Q} \parallel \mathbb{P}) \right\}, \quad X \in \mathcal{L}^2(\mathbb{P}), \label{MMV Definition}
\end{equation}
where $\mathbb{Q} \in \Delta^2(\mathbb{P}) \triangleq \left\{\mathbb{Q}:\mathbb{Q}(\Omega)=1, \mathbb{E}^\mathbb{P} \left[{\left(\frac{\mathrm{d}\mathbb{Q}}{\mathrm{d}\mathbb{P}}\right)}^2\right] < \infty \right\}$, and
\begin{equation}
    \mathbb{C}(\mathbb{Q} \parallel \mathbb{P}) \triangleq \left\{
    \begin{aligned}
    & \mathbb{E^\mathbb{P}}\left[{\left(\frac{\mathrm{d}\mathbb{Q}}{\mathrm{d}\mathbb{P}}\right)}^2 \right]-1, & \quad \text{if} \quad \mathbb{Q} \ll \mathbb{P}, \\
    & \infty, & \quad \text{otherwise},
    \end{aligned}
    \right. \notag
\end{equation}
is the relative Gini concentration index. 
According to \citet{maccheroni2009portfolio}, given the closed and convex domain of monotonicity of $U_{\gamma}$ defined by $\mathcal{G}_{\gamma} \triangleq \left\{X \in \mathcal{L}^2(\mathbb{P}): X - \mathbb{E}^\mathbb{P}\left[X\right] \leq \frac{1}{\gamma} \right\}$, $V_{\gamma}$ keeps the same value as $U_{\gamma}$ within $\mathcal{G}_{\gamma}$ and is the monotone envelope (the minimal dominating monotone function) of $U_{\gamma}$.

Naturally, we can rewrite the MMV preference as
\begin{equation}
	V_{\gamma}(X) = \inf_{Y \in \mathcal{M}} \mathbb{E}^\mathbb{P} \left[ XY + \frac{1}{2\gamma} Y^2 \right] - \frac{1}{2\gamma},    
\end{equation}
where $\mathcal{M} = \Big\{Y: Y \in \mathcal{L}^2(\mathbb{P}), \ \mathbb{E}^\mathbb{P}[Y]=1,\  Y \geq 0\Big\}$. It would be convenient for us to characterize ${V}_{\gamma}$ in the dynamic setting if the target $Y$ can be restricted on the set of all the non-negative Doléans-Dade exponentials. For a predictable process $\phi \triangleq \Big\{\phi(t) = \left(\phi_{1}(t), \phi_2(t,q) \right); t \in [0,T], q \in [-1, \infty) \Big\}$ and a process $Y^{\phi} \triangleq \Big\{Y^{\phi}(t); t \in [0,T]\Big\}$, we denote $Y^{\phi} = \mathcal{E}(\phi)$ if it solves SDE
\begin{equation}
    \mathrm{d}Y^{\phi}(t) = Y^{\phi}(t^{-}) \left(-\phi_1(t)\mathrm{d}B(t) - \int_{-1}^{\infty}\phi_2(t,q)\widetilde{N}(\mathrm{d}t,\mathrm{d}q) \right), \quad Y^{\phi}(0) = 1. \notag
\end{equation}
Denote $\varPhi = \Big\{\phi: \phi \ \text{is predictable}, \ \phi_2(t,q) \leq 1 \ \text{for all} \ (t,q) \in [0,T] \times [-1, \infty), Y^{\phi}(T) = \mathcal{E}(\phi)_T \in \mathcal{M} \Big\}$. We want to prove that the following associated preferences are equivalent, i.e., for all $X\in \mathcal{L}^2(\mathbb{P})$, the utility values under these four preferences are equal.
\begin{equation}
    \left\{
    \begin{aligned}
    V_{\gamma}(X) & = \inf_{Y \in \mathcal{M}} \mathbb{E}^\mathbb{P} \left[ XY + \frac{1}{2\gamma} Y^2\right] - \frac{1}{2\gamma}, \\
        V_{\gamma}^{1}(X) & = \inf_{Y \in \mathcal{M}_1} \mathbb{E}^\mathbb{P} \left[ XY + \frac{1}{2\gamma} Y^2\right] - \frac{1}{2\gamma}, \quad \mathcal{M}_1 = \Big\{ Y: Y \in \mathcal{M}, Y \geq \delta \ \text{for some} \ \delta > 0 \Big\}, \\
        V_{\gamma}^{2}(X) & = \inf_{Y \in \mathcal{M}_2} \mathbb{E}^\mathbb{P} \left[ XY + \frac{1}{2\gamma} Y^2\right] - \frac{1}{2\gamma}, \quad \mathcal{M}_2 = \Big\{ Y: Y \in \mathcal{M}, Y= \mathcal{E}(\phi)_T \ \text{for some} \ \phi \in \varPhi \Big\}, \\     
        V_{\gamma}^{3}(X) & = \inf_{Y \in \mathcal{M}_3} \mathbb{E}^\mathbb{P} \left[ XY + \frac{1}{2\gamma} Y^2\right] - \frac{1}{2\gamma}, \quad \mathcal{M}_3 = \mathcal{M}_1 \cap \mathcal{M}_2.
    \end{aligned}
    \right. \notag
\end{equation}
It is easy to see that $V_{\gamma}(X) \leq V_{\gamma}^{1}(X) \leq V_{\gamma}^{3}(X)$ and $V_{\gamma}(X) \leq V_{\gamma}^{2}(X) \leq V_{\gamma}^{3}(X)$. We only need to prove that $V_{\gamma}(X) =V_{\gamma}^{1}(X)$ and $V_{\gamma}^1(X) =V_{\gamma}^{3}(X)$. We have the following two lemmas and then the equivalent relationship between these MMV associated preferences.
\begin{lemma} \label{Lemma V and V1}
	For any $X \in \mathcal{L}^2(\mathbb{P})$, we have $V_{\gamma}(X)= V_{\gamma}^{1}(X)$.
\end{lemma}
\textbf{Proof.} 
For any $Y \in \mathcal{L}^2(\mathbb{P})$, if the sequence $\left\{Y_n\right\}_{n \geq 1}$ converges to $Y$ in $\mathcal{L}^2(\mathbb{P})$, then
\begin{equation}
    \lim_{n \to \infty} \{X Y_n + \frac{1}{2\gamma} Y_n^2 \}= X Y + \frac{1}{2\gamma} Y^2. \notag
\end{equation}
Thus, we only need to prove that $\mathcal{M}_1$ is dense in $\mathcal{M}$. For $Y \in \mathcal{M}$, we let
\begin{equation}
    Y_n = \max\left\{Y,\frac{1}{n}\right\}, \quad Y_n^{\prime}=\frac{Y_n}{\mathbb{E}\left[Y_n\right]}. \notag
\end{equation}
Then, $Y_n^{\prime} \in \mathcal{M}_1$. Moreover, we have $Y_n \geq Y$, $\mathbb{E}^\mathbb{P} \left[ Y_n \right] \geq \mathbb{E}^\mathbb{P} [Y] = 1$ and then $0\leq Y_n'\leq Y_n\leq Y+1$. Using  Dominated convergence theorem, we have $Y_n^{\prime} \to Y$ in $\mathcal{L}^2(\mathbb{P})$, which completes the proof. \hfill $\square$ 
\vskip 5pt

\begin{lemma} \label{Lemma V1 and V3}
	For any $X \in \mathcal{L}^2(\mathbb{P})$, we have $V_{\gamma}^{1}(X)= V_{\gamma}^{3}(X)$.
\end{lemma}
\textbf{Proof.} 
Using the martingale representation theorem (see Chapter 9 of  \citet{tankov2003jumpmodeling}), for any $Y \in \mathcal{M}_1$, there exists a predictable $\phi$ such that 
\begin{equation}
    Y = 1 - \int_{0}^{T}\phi_{1}(t) \mathrm{d} B(t)-\int_{0}^{T}\int_{-1}^{\infty} \phi_2(t,q) \widetilde{N}(\mathrm{d}t, \mathrm{d}q). \notag
\end{equation}
As $Y(t)=\mathbb{E}^\mathbb{P} \left[Y \mid \F_t\right]\geq \delta$ and $Y(t^-)=\lim_{s \to t^-}Y(s^-) \geq \delta$, we let 
\begin{equation}
    \phi_1^{\prime}(t) = \frac{\phi_1(t)}{Y(t^-)}, \quad \phi_2^{\prime}(t,q)=\min\left\{\frac{\phi_2(t,q)}{Y(t^-)},1\right\}, \notag
\end{equation}
and then 
\begin{equation}
    \mathrm{d} Y(t) = Y(t^-) \phi_{1}^{\prime} (t) \mathrm{d} B(t) - \int_{-1}^{\infty}Y(t^-) \phi_{2}^{\prime} (t,q) \widetilde{N}(\mathrm{d}t, \mathrm{d}q). \notag
\end{equation}
Besides, $Y(t) \geq 0$ indicates that $\phi_2(t,\Delta L(t))\one_{\Delta L(t) \neq 0} = - \Delta Y(t) \leq Y(t^-)$. Then, $\phi_2^{\prime}(t, \Delta L(t)) \one_{\Delta L(t) \neq 0} \leq 1$ and we can always choose $\phi_2^{\prime}$ such that  $\phi_2^{\prime} (t,q) \leq 1$ for any $(t, q) \in [0,T] \times [-1,\infty)$. Thus, $Y= \mathcal{E}(\phi)_T \ \text{for some} \ \phi \in \varPhi$, which means  $V_{\gamma}^{1}(X)= V_{\gamma}^{3}(X)$. \hfill $\square$
\vskip 5pt

\begin{theorem} \label{Theorem MMV V2 Representation}
	For any $X \in \mathcal{L}^2(\mathbb{P})$, we have $V_{\gamma}(X) = V_{\gamma}^{1}(X) = V_{\gamma}^{2}(X) = V_{\gamma}^{3}(X)$ and 
    \begin{equation}
        V_\gamma(X) = \inf_{Y = \mathcal{E}(\phi)_T, \phi \in \varPhi} \mathbb{E}^\mathbb{P} \left[ XY + \frac{1}{2\gamma} Y^2\right] - \frac{1}{2\gamma}. \label{MMV V2 Representation}
    \end{equation}
\end{theorem}
\textbf{Proof.} Using Lemmas \ref{Lemma V and V1} and \ref{Lemma V1 and V3}, we have $V_{\gamma}(X) = V_{\gamma}^{1}(X) = V_{\gamma}^{3}(X)$. Then, due to the relationship $ V_{\gamma}(X) \leq V_{\gamma}^{2}(X) \leq V_{\gamma}^{3}(X)$, we conclude that $V_{\gamma}(X) = V_{\gamma}^{1}(X) = V_{\gamma}^{2}(X) = V_{\gamma}^{3}(X)$ and \eqref{MMV V2 Representation} is given by $V_{\gamma}(X) = V_{\gamma}^{2}(X)$. \hfill $\square$ 
\vskip 5pt

Theorem \ref{Theorem MMV V2 Representation} is significant for us to characterize the dynamic of process $Y$. Now, we can redefine $Y^{\phi} \triangleq \Big\{Y^{\phi}(s); s \in [t,T]\Big\}$ by
\begin{equation}
    \mathrm{d}Y^{\phi}(s) = Y^{\phi}(s^{-}) \left(-\phi_1(s)\mathrm{d}B(s) - \int_{-1}^{\infty}\phi_2(s,q)\widetilde{N}(\mathrm{d}s,\mathrm{d}q) \right), \quad Y^{\phi}(t) = y = \frac{1}{2\gamma}. \label{Y Process}
\end{equation}
Consider the probability measure $\mathbb{Q}^{\phi}$ generated by $\frac{\mathrm{d}\mathbb{Q}^{\phi}}{\mathrm{d}\mathbb{P}} = \frac{Y^{\phi}(T)}{y}$. By using the Girsanov transformation, the process $B^{\phi}$ defined by 
\begin{equation}
    \mathrm{d}B^{\phi}(s) \triangleq \mathrm{d}B(s) + \phi_1(s)\mathrm{d}s, \quad s\in[t,T],\notag
\end{equation}
is a standard $\mathbb{Q}^{\phi}$-Brownian motion, and the process $\widetilde{N}^{\phi}$ defined by
\begin{equation}
    \widetilde{N}^{\phi}(\mathrm{d}s,\mathrm{d}q) \triangleq \widetilde{N}(\mathrm{d}s,\mathrm{d}q) + \phi_2 (s,q) \nu (\mathrm{d}q)\mathrm{d}s, \quad s \in [t,T],\notag
\end{equation}
is a $\mathbb{Q}^{\phi}$-compensated Poisson random measure with the compensator $\left(1-\phi_2(t,q)\right) \nu (\mathrm{d}q)$. Then, the investor's wealth process $X^\pi$ and the Doléans-Dade exponential process $Y^{\phi}$ follow a dynamic system:
\begin{equation}
    \left\{
    \begin{aligned}
    \mathrm{d} X^{\pi}\left(s\right) & = \Bigg[ X^{\pi}\left(s\right)r +  \pi \left(s\right)\left(\mu-r + \lambda \xi_1 - \sigma \phi_1\left(s\right) - \int_{-1}^{\infty}q\phi_2(s,q)\nu \left(\mathrm{d}q\right) \right) \Bigg]\mathrm{d}s  \\
    & + \pi \left(s\right) \sigma \mathrm{d} B^{\phi} \left(s\right) + \pi \left(s^-\right)\int_{-1}^{\infty} q \widetilde{N}^{\phi} \left(\mathrm{d}s,\mathrm{d}q\right), \\
    \mathrm{d} Y^{\phi} \left(s\right) & = Y^{\phi} \left(s^{-}\right)  \Bigg( {\phi_1 \left(s\right)}^2 + \int_{-1}^{\infty} {\phi_2 \left(s,q\right)}^2 \nu \left(\mathrm{d}q \right) \Bigg) \mathrm{d}s -  Y^{\phi} \left(s^{-}\right) \phi_1 \left(s\right) \mathrm{d}B^{\phi} \left(s\right) \\
    & - Y^{\phi} \left(s^{-}\right) \int_{-1}^{\infty} \phi_2 \left(s,q\right) \widetilde{N}^{\phi} \left( \mathrm{d}s,\mathrm{d}q \right) \\
    \end{aligned}
    \right. \label{Dynamicsystem} \notag
\end{equation}
with the initial condition $X^{\pi}(t) = x, Y^{\phi}(t) = y$. We temporarily allow $y$ to be any non-negative value, as we will take the dynamic programming method to solve the control problem. Based on Theorem \ref{Theorem MMV V2 Representation}, we represent the MMV preference \eqref{MMV Definition} by
\begin{equation}
    V_{\gamma}(X) = \inf_{\phi \in \varPhi}  \mathbb{E}^{\mathbb{Q}^{\phi}} \left\{ X + \frac{1}{2\gamma} \frac{\mathrm{d}\mathbb{Q}^{\phi}}{\mathrm{d}\mathbb{P}} \right\} - \frac{1}{2\gamma}, \quad X \in \mathcal{L}^2(\mathbb{P}). \label{MMV Auxiliary Preference}
\end{equation}

\begin{definition}
    Suppose that the market player's initial state is $y$ at time $t$. A strategy $\phi \triangleq \Big\{\phi(s) = \left(\phi_1(s),\phi_2(s,q)\right); s \in [t,T], q \in [-1, \infty) \Big\}$ is admissible and denoted as $\phi \in \varPhi_t$ if it satisfies:
    \begin{enumerate}
        \item $\phi$ is predictable with respect to the filtration $\left\{  \mathcal{F}_{s}\right\}_{s\in[t,T]}$.
        \item For any $s \in [t,T], q \in [-1, \infty)$, we have $\phi(s, q) \leq 1$.
        \item SDE \eqref{Y Process} has a unique solution which is a non-negative square-integrable martingale.
    \end{enumerate}
\end{definition}

\begin{definition}
    Suppose that the investor's initial wealth level is $x$ at time $t$. A trading strategy $\pi \triangleq \Big\{\pi(s); s \in [t,T] \Big\}$ is admissible and denoted as $\pi \in \mathcal{A}^{mmv}_t$ if it satisfies the following conditions:
    \begin{enumerate}
        \item $\pi$ is predictable with respect to the filtration $\left\{  \mathcal{F}_{s}\right\}_{s\in[t,T]}$.
        \item For any $\phi \in \varPhi_t$, we have  $\mathbb{E}_{t,x,y}^{\mathbb{Q}^{\phi}}\left[\sup_{t \leq s \leq T} {\left| X^{\pi}(s) \right|}^2 \right] < \infty$.
        \item SDE \eqref{Jump-Diffusion Wealth Process} has a unique solution.
    \end{enumerate}
\end{definition}
\vskip 5pt
Based on admissible sets defined above and varying the initial value $y$, we define the following max-min investment problem. 

\begin{problem} \label{Problem MMV Auxiliary}
    Under the auxiliary preference \eqref{MMV Auxiliary Preference}, the investor's optimization goal is to solve
    \begin{equation}
         \sup_{\pi \in \mathcal{A}^{mmv}_t} \inf_{\phi \in \varPhi_t} J\left(t,x,y,\pi,\phi \right) \triangleq \sup_{\pi \in \mathcal{A}^{mmv}_t} \inf_{\phi \in \varPhi_t} \mathbb{E}_{t,x,y}^{\mathbb{Q}^{\phi}} \left[ X^{\pi} \left(T\right) + Y^{\phi} \left(T\right) \right], \label{MMV Auxiliary Function}
    \end{equation}
    When we set $y=\frac{1}{2\gamma}$, the above problem becomes the MMV investment problem. This problem is a robust control problem and can be regarded as a zero-sum stochastic differential game between the investor and the market player. We hope to find a saddle point $(\pi^{\ast},\phi^{\ast}) \in \mathcal{A}^{mmv}_t\times \varPhi_t$ such that
    \begin{equation}
         J\left(t,x,y,\pi,\phi^{\ast} \right) \leq J\left(t,x,y,\pi^{\ast},\phi^{\ast} \right) \leq J\left(t,x,y,\pi^{\ast},\phi \right), \  \text{for all} \ (\pi,\phi) \in \mathcal{A}^{mmv}_t\times \varPhi_t, \notag
    \end{equation}
    and a value function $W(\cdot,\cdot,\cdot)$ satisfying
    \begin{equation}
        \left\{
        \begin{aligned}
            & W(t,x,y) = \sup_{\pi \in \mathcal{A}^{mmv}_t} \inf_{\phi \in \varPhi_t} \mathbb{E}_{t,x,y}^{\mathbb{Q}^{\phi}} \Big[ X^{\pi} \left(T\right) + Y^{\phi} \left(T\right) \Big], \\
            & W(T,x,y) = x + y. \label{MMV Auxiliary Objective} \notag
        \end{aligned}
        \right. 
    \end{equation}
\end{problem}
\vskip 5pt
\subsection{Verification Theorem and Solution}
\noindent
Problem \ref{Problem MMV Auxiliary} is a robust control problem and can be dealt with the HJBI equation. Referring to \citet{mataramvura2008risk} and \citet{trybula2019continuous}, we establish a verification theorem for this problem. For any $W\left(\cdot,\cdot,\cdot \right) \in \mathcal{C}^{1,2,2} \left( \left[0,T\right) \times \mathbb{R} \times \left(0,+ \infty \right)\right)$, define the differential operator $
\varPsi^{a,\phi}$ as
\begin{align}
    \varPsi^{\pi,\phi}W\left(t,x,y\right) & = W_t + \Bigg[xr + \pi\left(t\right) \left(\mu - r + \lambda \xi_1 -\sigma \phi_1 \left(t \right) - \int_{-1}^{\infty}q\phi_2(t,q)\nu \left(\mathrm{d}q\right)  \right) \Bigg] W_x \notag \\
    & + y \Bigg( {\phi_1\left(t\right)}^2 + \int_{-1}^{\infty}{\phi_2 \left(t,q\right)}^2 \nu \left( \mathrm{d}q \right) \Bigg) W_y + \frac{1}{2} {\pi \left(t\right)}^2 \sigma^2 W_{xx} + \frac{1}{2} y^2 {\phi_1 \left(t\right)}^2 W_{yy} \notag \\
    & - \pi \left(t\right) \sigma y \phi_1 \left(t\right)W_{xy} + \int_{-1}^{\infty} \Big[ W\left(t,x+\pi\left(t\right)q, y - y \phi_2 \left(t,q\right) \right) - W \left(t,x,y\right) \notag \\
    & - \pi\left(t\right)q W_{x} + y \phi_2 \left(t,q\right) W_y \Big] \left(1-\phi_2\left(t,q\right)\right) \nu \left(\mathrm{d} q \right). \notag
\end{align}

\begin{theorem}\label{Theorem MMV Verification}
    Suppose that there exist a real value function $W\left(t,x,y\right) \in \mathcal{C}^{1,2,2} \left( \left[0,T\right) \times \mathbb{R} \times \mathbb{R}^+\right)$ and a control $\left( \pi^{\ast}, \phi^{\ast} \right) \in \mathcal{A}^{mmv}_t \times \varPhi_t$ satisfying the following conditions:
    \begin{enumerate}
        \item For any $(\pi,\phi) \in \mathcal{A}^{mmv}_t \times \varPhi_t$ and $\left(t,x,y\right) \in  \left[0,T\right) \times \mathbb{R} \times \mathbb{R}^+$,
        \begin{equation}
            \varPsi^{\pi^{\ast},\phi}W\left(t,x,y\right) \geq 0, \quad \varPsi^{\pi,\phi^{\ast}}W\left(t,x,y\right) \leq 0, \quad       \varPsi^{\pi^{\ast},\phi^{\ast}}W\left(t,x,y\right) = 0, \quad W\left(T,x,y\right)=x+y. \notag
        \end{equation}
        \item For any $(\pi,\phi) \in \mathcal{A}^{mmv}_t \times \varPhi_t$ and $\left(t,x,y\right) \in  \left[0,T\right) \times \mathbb{R} \times \mathbb{R}^+$,
        \begin{equation}
            \mathbb{E}_{t,x,y}^{\mathbb{Q}^{\phi}} \left[ \sup_{t \leq s \leq T} \left| W \left(s, X^{\pi}\left(s\right) , Y^{\phi}\left(s\right) \right)  \right| \right] < \infty. \label{Condition_5} \notag
        \end{equation}
    \end{enumerate}
    Then $W(t,x,y)$ is the value function of the robust control Problem \ref{Problem MMV Auxiliary} and $\left(\pi^{\ast}, \phi^{\ast} \right)$ is the corresponding optimal control, i.e.,
    \begin{align}
    W\left(t,x,y\right) & = J\left(t,x,y,\pi^{\ast},\phi^{\ast}\right) = \sup_{\pi \in \mathcal{A}^{mmv}_t} J\left(t,x,y,\pi,\phi^{\ast} \right)  = \inf_{\phi \in \varPhi_t}J\left(t,x,y,\pi^{\ast},\phi \right)  \notag \\
    & = \sup_{\pi \in \mathcal{A}^{mmv}_t}  \inf_{\phi \in \varPhi_t} J\left(t,x,y,\pi,\phi \right)  = \inf_{\phi \in \varPhi_t}  \sup_{\pi \in \mathcal{A}^{mmv}_t} J\left(t,x,y,\pi,\phi \right). \notag
    \end{align} 
\end{theorem}
\textbf{Proof.} See Appendix \ref{Appendix Proof MMV Verification} for the detailed proof. The key difference between our verification theorem and that of \citet{trybula2019continuous} is that we restrict our target optimal control in the admissible set $\mathcal{A}^{mmv}_t \times \varPhi_t$. Using such replacement, we prove that the verification theorem is still valid and we can solve out the optimal control in $\mathcal{A}^{mmv}_t \times \varPhi_t$ explicitly. \hfill $\square$ 
\vskip 5pt
To find out the explicit form of value function $W(t,x,y)$ and the saddle point $\left(\pi^{\ast}, \phi^{\ast} \right)$ of Problem \ref{Problem MMV Auxiliary}, we start with analyzing the HJBI equation with the boundary condition:
\begin{equation}
    \left\{
    \begin{aligned}
        & \sup_{\pi \in \mathbb{R}} \inf_{\phi_1 \in\mathbb{R}, \phi_2 \leq 1} \varPsi^{\pi,\phi}W \left(t,x,y\right)=0, \\
        & W(T,x,y) = x + y. \label{HJBI Equation}
    \end{aligned}
    \right.
\end{equation}

We guess that the solution has the following linear form: 
\begin{equation}
    W\left(t,x,y\right) = A\left(t\right) x + B\left(t\right)y, \quad \text{with} \quad A(T) = 1, \quad B(T) = 1. \notag
\end{equation}
Substituting it into the HJBI equation \eqref{HJBI Equation}, we have
\begin{align}\label{hjb1}
    \dot{A}(t)x + \dot{B}(t)y & + \sup_{\pi} \inf_{\phi} \Bigg\{ \Bigg[xr + \pi \left(t\right) \left(\mu - r + \lambda \xi_1 -\sigma \phi_1 \left(t\right) - \int_{-1}^{\infty}q\phi_2(t,q)\nu \left(\mathrm{d}q\right) \right) \Bigg] A(t) \notag \\
    & + y \Bigg( {\phi_1\left(t\right)}^2 + \int_{-1}^{\infty}{\phi_2 \left(t,q\right)}^2 \nu(\mathrm{d}q)\Bigg) B(t) \Bigg\} = 0. 
\end{align}
Assume  $A(t) \geq 0$, $B(t) \geq 0$. By the first-order condition, the minimization over $\phi \in \varPhi_t$ is attained at:
\begin{equation}
    \phi_1 ^{0} \left(\pi \right)=\frac{\pi\left(t\right)\sigma A\left(t\right)}{2yB\left(t\right)}, \quad \phi_2^{0}\left(\pi,q\right) = \min\left\{\frac{\pi(t)q A(t)}{2y B(t)}, 1 \right\}. \notag
\end{equation}
As $-1 \leq q < \infty$, there exists $q^{mmv}(t)$ such that $\frac{\pi(t)q^{mmv}(t) A(t)}{2y B(t)} = 1$, i.e., $q^{mmv}(t) = \frac{2y B(t)}{\pi(t)A(t)}$.  We temporarily write $\phi_2^{0}\left(\pi,q\right) = \min\left\{\frac{q}{q^{mmv}(t)}, 1 \right\}$. Substituting it into Equation \eqref{hjb1}, the maximization over $\pi \in \mathbb{R}$ is attained at:
\begin{equation}
    \pi ^{0} \left(t\right) = \frac{2y\left(\mu-r +\lambda\xi_1 - \int_{-1}^{\infty} \widetilde{q} \min\left\{\frac{\widetilde{q}}{q^{mmv}(t)}, 1 \right\}\nu(\mathrm{d}\widetilde{q}) \right)B\left(t\right)}{\sigma^2 A\left(t\right)}. \notag
\end{equation}
We should have $q^{mmv}(t)= \frac{2y B(t)}{\pi^0(t)A(t)}$, which indicates that $q^{mmv}(t)$ satisfies
\begin{equation}
    \mu-r +\lambda\xi_1 - \int_{-1}^{\infty} \widetilde{q} \min\left\{\frac{\widetilde{q}}{q^{mmv}(t)}, 1 \right\}\nu(\mathrm{d}\widetilde{q})=\sigma^2\frac{1}{q^{mmv}(t)}. \notag
\end{equation}
Next, we prove that the above function has a unique solution on $[-1,\infty)$.
\begin{lemma} \label{Lemma Q Existence and Uniqueness}
    The function 
    \begin{equation}
        f(x) =\sigma^2-x \left(\mu-r +\lambda\xi_1 - \int_{-1}^{\infty} \widetilde{q} \min\left\{\frac{\widetilde{q}}{x}, 1 \right\}\nu(\mathrm{d}\widetilde{q})\right),
    \end{equation}
    has a unique zero point on $[-1, \infty)$, denoted as $q^{mmv}$. Furthermore, $q^{mmv}\in(0, \infty)$, and it satisfies 
    \begin{equation}
        \frac{\sigma^2 + \int_{-1}^{q^{mmv}} \widetilde{q}^2 \nu(\mathrm{d}\widetilde{q})}{\mu-r + \int_{-1}^{q^{mmv}} \widetilde{q} \nu(\mathrm{d}\widetilde{q})}=q^{mmv}. \label{Q Equation}
    \end{equation}
\end{lemma}
\textbf{Proof.} See Appendix \ref{Appendix Proof Q Existence and Uniqueness} for the detailed proof.	\hfill $\square$ 
\vskip 5pt
Now, the corresponding $\phi^{\ast}$ is
\begin{equation}
\begin{aligned}
    \big(\phi_1 ^{\ast}, \phi_2^{\ast} \left(q\right) \big) =& \left(\frac{\left(\mu-r + \int_{-1}^{q^{mmv}} \widetilde{q} \nu(\mathrm{d}\widetilde{q})\right)\sigma}{\sigma^2 + \int_{-1}^{q^{mmv}} \widetilde{q}^2 \nu(\mathrm{d}\widetilde{q})}, \min\left\{ \frac{\left(\mu-r + \int_{-1}^{q^{mmv}} \widetilde{q} \nu(\mathrm{d}\widetilde{q})\right)q}{\sigma^2 + \int_{-1}^{q^{mmv}} \widetilde{q}^2 \nu(\mathrm{d}\widetilde{q})}, 1 \right\}\right) \\=& \left(\frac{\sigma}{q^{mmv}}, \min \left\{\frac{q}{q^{mmv}}, 1 \right\}\right), \notag
\end{aligned}
\end{equation}
then the HJBI equation becomes
\begin{equation}
    \left[\dot{A}\left(t\right)+rA\left(t\right)\right]x+\left[\dot{B}\left(t\right)+C^{mmv} B\left(t\right)\right]y= 0, \notag
\end{equation}
where $C^{mmv} = \frac{\left(\mu-r + \int_{-1}^{q^{mmv}} \widetilde{q} \nu(\mathrm{d}\widetilde{q})\right)^{2}}{\sigma^2 + \int_{-1}^{q^{mmv}} \widetilde{q}^2 \nu(\mathrm{d}\widetilde{q})} + \int_{q^{mmv+}}^{\infty} \nu(\mathrm{d}\widetilde{q})$ is a constant. As $x$ and $y$ are arbitrary, we have:
\begin{equation}
    \left\{
    \begin{aligned}
    & \dot{A}\left(t\right)+rA\left(t\right)=0, \quad A(T) = 1, \\
    & \dot{B}\left(t\right)+C^{mmv}B\left(t\right)=0, \quad B(T)=1.
    \end{aligned}
    \right. \notag
\end{equation}
Solving ODEs, we have $A\left(t\right)=e^{\left(T-t\right)r}, B\left(t\right)=e^{\left(T-t\right)C^{mmv}}$, consistent with the assumption $A(t) \geq 0$, $B(t) \geq 0$. Next, we provide a theorem to verify the admissibility and optimality of $(\pi^{\ast}, \phi^{\ast})$.

\begin{theorem}\label{Theorem MMV Function and Control}
For Problem \ref{Problem MMV Auxiliary}, the value function is given by
\begin{equation}
    W(t,x,y) = e^{\left(T-t\right)r}x + e^{\left(T-t\right)C^{mmv}}y, \notag
\end{equation}
and the optimal control is
\begin{equation}
    \left\{
    \begin{aligned}
        \pi^{\ast}(s) & = \frac{2}{q^{mmv}} e^{(T-s)(C^{mmv}-r)} Y^{\phi^{\ast}}(s^{-}),  \\
        \phi^{\ast}(s) & = \big(\phi_1^{\ast}, \phi_2^{\ast}(q) \big) = \left(\frac{\sigma}{q^{mmv}}, \min \left\{\frac{q}{q^{mmv}}, 1\right\}\right), 
    \end{aligned}
    \right. \notag
\end{equation}
for any $s \in [t, T]$ and $q \in [-1, \infty)$, where $C^{mmv} = \frac{\left(\mu-r + \int_{-1}^{q^{mmv}} \widetilde{q} \nu(\mathrm{d}\widetilde{q})\right)^{2}}{\sigma^2 + \int_{-1}^{q^{mmv}} \widetilde{q}^2 \nu(\mathrm{d}\widetilde{q})} + \int_{q^{mmv+}}^{\infty} \nu(\mathrm{d}\widetilde{q})$, $q^{mmv}$ is the solution of Equation \eqref{Q Equation}, and $Y^{\phi^{\ast}}$ is the process satisfying \eqref{Y Process} with $\phi^{\ast}$. 
\end{theorem}
\textbf{Proof.} See Appendix \ref{Appendix Proof MMV Function and Control} for the detailed proof. Besides, the proof also derives out Theorem \ref{Theorem MMV investor}, Proposition \ref{Proposition MMV Stopping Time} and Proposition \ref{Proposition MMV Expression}. 
\hfill $\square$ 
\vskip 5pt

Based on Theorem \ref{Theorem MMV Function and Control}, setting the initial value $y = \frac{1}{2\gamma}$, we can solve the original MMV problem. 

\begin{theorem}\label{Theorem MMV investor}
    For an investor with initial wealth $x$ at time $t$, the optimal investment strategy under MMV preferences is 
    \begin{equation}
        \pi^{mmv,\ast}(s) = \frac{1}{q^{mmv}} \max\left\{ e^{(s-t)r}x + \frac{e^{(T-t)C^{mmv}-(T-s)r}}{\gamma} -X^{mmv,\ast}(s^-), 0\right\} \label{MMV Strategy}
    \end{equation}
     for any $s \in [t, T]$, where $q^{mmv}$ is the solution of Equation \eqref{Q Equation}, $C^{mmv} = \frac{\left(\mu-r + \int_{-1}^{q^{mmv}} \widetilde{q} \nu(\mathrm{d}\widetilde{q})\right)^{2}}{\sigma^2 + \int_{-1}^{q^{mmv}} \widetilde{q}^2 \nu(\mathrm{d}\widetilde{q})} + \int_{q^{mmv+}}^{\infty} \nu(\mathrm{d}\widetilde{q})$, and $X^{mmv,\ast}$ denotes the wealth process under the strategy $\pi^{mmv,\ast}$. The maximum utility gained by the MMV investor is
    \begin{equation}
        V_\gamma\left(X^{mmv,\ast}(T)\right) = e^{(T-t)r}x + \frac{e^{(T-t)C^{mmv}}-1}{2\gamma}. 
    \end{equation}
\end{theorem}

Similarly, we give the following two propositions based on Theorem \ref{Theorem MMV investor} to better understand the MMV investor's behavior. 

\begin{proposition} \label{Proposition MMV Stopping Time}
    Define a stopping time $T_{q^{mmv}} \triangleq \inf \Big\{s > t: \Delta{L}(s) \geq q^{mmv} \Big\}$, and $\inf\{ \varnothing\}=\infty$. When $t \leq s \leq T_{q^{mmv}}$, we have $\pi^{mmv,\ast}(s) > 0$, which means that the investor holds a positive amount of the risky asset. When $ s > T_{q^{mmv}}$, we have $\pi^{mmv,\ast}(s) = 0$, which means that  the investor changes to keep zero position of the risky asset after the relatively large jump occurs.  
\end{proposition}

Proposition \ref{Proposition MMV Stopping Time} shows that MMV investor tends to hold zero position of the risky asset forever when $\Delta{L} > q^{mmv}$ happens, which is essentially different from the MV investor in Proposition \ref{Proposition MV Stopping Time}. For such phenomenon, more explanations and visualized examples are provided in Section \ref{Section Numerical}.

\begin{proposition} \label{Proposition MMV Expression}
    When $t \leq s \leq T$, the explicit form of the wealth process can be given by
    \begin{align}
        X^{mmv,\ast}(s) & = e^{(s-t)r}x + \frac{1}{\gamma}e^{(T-t)C^{mmv}-(T-s)r} - 2e^{(T-s)(C^{mmv}-r)}Y^{\phi^{\ast}}(s) \one_{\{t \leq s < T_{q^{mmv}}\}} \notag \\
        & + 2e^{(T-T_{q^{mmv}})C^{mmv}- (T-s)r} \left(\frac{\Delta L(T_{q^{mmv}})}{q^{mmv}} - 1 \right) Y^{\phi^{\ast}}(T_{q^{mmv}}^-) \one_{\{s \geq T_{q^{mmv}}\}}, \notag 
    \end{align}
    where $Y^{\phi^{\ast}}(s)$ is the Doléans-Dade exponential
    \begin{equation}
    Y^{\phi^{\ast}}(s) = \frac{1}{2\gamma} \mathcal{E}\Bigg(-\int_t^. \phi_1^{\ast} \mathrm{d}B(\tau) - \int_t^. \int_{-1}^{\infty} \phi_2^{\ast}(q) \widetilde{N} \left(\mathrm{d}\tau,\mathrm{d}q \right) \Bigg)_s. \notag
    \end{equation}
\end{proposition}

\section{Economic Essence of MMV Preferences}{\label{Section Consistency}}
\noindent
In this section, we begin by proposing a sufficient and necessary condition for solutions to MV and MMV problems to be consistent, which also indicates that they can be different in the jump-diffusion model. Then, we provide an explanation based on the pricing operator for the inconsistency due to the jump. As such, we reveal the essential difference between MV and MMV, and show that MMV should be more economically reasonable due to its non-monotonicity.

\subsection{Comparison of MV and MMV Investment Behaviors}{\label{Subsection Consistency}}
\noindent
The results given in Theorems \ref{Theorem MV Investor} and  \ref{Theorem MMV investor} look similar but with notable differences.   First, the coefficient term $\frac{1}{q^{mv}}$ is different from the term $\frac{1}{q^{mmv}}$. Second, the exponential term $C^{mv}$ is different from the term $C^{mmv}$. Such differences lead to different optimal strategies and maximum utility values under MV and MMV. Next, we study the relationship between them. 

\begin{lemma} \label{Lemma Q Equivalence}
    We have $q^{mmv} \leq q^{mv}$, where $q^{mmv}$ is the solution of Equation: $q^{mmv} = \frac{\sigma^2 + \int_{-1}^{q^{mmv}} \widetilde{q}^2 \nu(\mathrm{d}\widetilde{q})}{\mu-r + \int_{-1}^{q^{mmv}} \widetilde{q} \nu(\mathrm{d}\widetilde{q})}$, and $q^{mv} \triangleq \frac{\sigma^2 + \lambda \xi_2^2}{\mu - r + \lambda \xi_1}$. Further, $q^{mmv} = q^{mv}$ if and only if $\Delta{L} \leq q^{mv}$ a.s..
\end{lemma}
\textbf{Proof.} 
Based on the definition of $\lambda \xi_2^2$ and $\lambda \xi_1$, we have
\begin{equation}
    q^{mmv} = \frac{\sigma^2 + \int_{-1}^{q^{mmv}} \widetilde{q}^2 \nu(\mathrm{d}\widetilde{q})}{\mu-r + \int_{-1}^{q^{mmv}} \widetilde{q} \nu(\mathrm{d}\widetilde{q})} =  \frac{\sigma^2 + \lambda \xi_2^2 - \int_{q^{mmv+}}^{\infty} \widetilde{q}^2 \nu(\mathrm{d}\widetilde{q})}{\mu-r + \lambda\xi_1 - \int_{q^{mmv+}}^{\infty} \widetilde{q} \nu(\mathrm{d}\widetilde{q})} \leq \frac{\sigma^2 + \lambda \xi_2^2 - q^{mmv}\int_{q^{mmv+}}^{\infty} \widetilde{q} \nu(\mathrm{d}\widetilde{q})}{\mu-r + \lambda\xi_1 - \int_{q^{mmv+}}^{\infty} \widetilde{q} \nu(\mathrm{d}\widetilde{q})}. \notag
\end{equation}
Simple calculation yields $q^{mmv} = \frac{\sigma^2 + \lambda \xi_2^2}{\mu - r + \lambda \xi_1} \leq q^{mv}$. And the equality holds if and only if $\int_{q^{mmv+}}^{\infty} \widetilde{q}^2 \nu(\mathrm{d}\widetilde{q}) = q^{mmv} \int_{q^{mmv+}}^{\infty} \widetilde{q} \nu(\mathrm{d}\widetilde{q})$, i.e., $\Delta{L} \leq q^{mv}$ a.s.. \hfill $\square$ 
\vskip 5pt

\begin{lemma} \label{Lemma C and C' Relationship}
    We have $C^{mmv} \geq C^{mv}$, where $C^{mmv} = \frac{\left(\mu-r + \int_{-1}^{q^{mmv}} \widetilde{q} \nu(\mathrm{d}\widetilde{q})\right)^{2}}{\sigma^2 + \int_{-1}^{q^{mmv}} \widetilde{q}^2 \nu(\mathrm{d}\widetilde{q})} + \int_{q^{mmv+}}^{\infty} \nu(\mathrm{d}\widetilde{q})$ and $C^{mv} = \frac{(\mu-r+\lambda \xi_1)^2}{\sigma^2 + \lambda \xi_2^2}$. Further, $C^{mmv} = C^{mv}$ if and only if $\Delta{L} \leq q^{mv}$ a.s.. 
\end{lemma}
\textbf{Proof.} 
Based on the definition of $q^{mmv}$, we have 
\begin{equation}
    \mu-r = \frac{\sigma^2 + \int_{-1}^{q^{mmv}} \widetilde{q}^2 \nu(\mathrm{d}\widetilde{q})}{q^{mmv}} - \int_{-1}^{q^{mmv}} \widetilde{q} \nu(\mathrm{d}\widetilde{q}). \notag
\end{equation}
Then, $C^{mmv} \geq C^{mv}$ is equivalent to 
\begin{equation}
    \frac{\sigma^2}{q^{mmv2}} + \frac{1}{q^{mmv2}} \int_{-1}^{q^{mmv}} \widetilde{q}^2 \nu(\mathrm{d}\widetilde{q}) + \int_{q^{mmv+}}^{\infty} \nu(\mathrm{d}\widetilde{q}) \geq \frac{\left(\frac{\sigma^2}{q^{mmv}} + \frac{1}{q^{mmv}} \int_{-1}^{q^{mmv}} \widetilde{q}^2 \nu(\mathrm{d}\widetilde{q}) + \int_{q^{mmv+}}^{\infty} \widetilde{q} \nu(\mathrm{d}\widetilde{q}) \right)^2}{\sigma^2 +  \int_{-1}^{q^{mmv}} \widetilde{q}^2 \nu(\mathrm{d}\widetilde{q}) + \int_{q^{mmv+}}^{\infty} \widetilde{q}^2 \nu(\mathrm{d}\widetilde{q})}. \notag
\end{equation}
Using Cauchy's inequality twice, we have 
\begin{align}
    & \left(\frac{\sigma^2}{q^{mmv2}} + \frac{1}{q^{mmv2}} \int_{-1}^{q^{mmv}} \widetilde{q}^2 \nu(\mathrm{d}\widetilde{q}) + \int_{q^{mmv+}}^{\infty} \nu(\mathrm{d}\widetilde{q})\right) \left(\sigma^2 +  \int_{-1}^{q^{mmv}} \widetilde{q}^2 \nu(\mathrm{d}\widetilde{q}) + \int_{q^{mmv+}}^{\infty} \widetilde{q}^2 \nu(\mathrm{d}\widetilde{q})\right) \notag \\
    \geq & \left(\frac{\sigma^2}{q^{mmv}} + \frac{1}{q^{mmv}} \int_{-1}^{q^{mmv}} \widetilde{q}^2 \nu(\mathrm{d}\widetilde{q}) + \sqrt{\int_{q^{mmv+}}^{\infty} \nu(\mathrm{d}\widetilde{q})\int_{q^{mmv+}}^{\infty} \widetilde{q}^2 \nu(\mathrm{d}\widetilde{q})} \right)^2 \notag \\
    \geq & \left(\frac{\sigma^2}{q^{mmv}} + \frac{1}{q^{mmv}} \int_{-1}^{q^{mmv}} \widetilde{q}^2 \nu(\mathrm{d}\widetilde{q}) + \int_{q^{mmv+}}^{\infty} \widetilde{q} \nu(\mathrm{d}\widetilde{q}) \right)^2. \notag
\end{align} Thus, $C^{mmv} \geq C^{mv}$ and the equality holds if and only if $\nu\left((q^{mmv},\infty)\right) = 0$, i.e., $\Delta{L} \leq q^{mv}$ a.s.. \hfill $\square$ 
\vskip 5pt

Based on the two lemmas, we provide the sufficient and necessary condition for MV and MMV to be consistent.
\begin{theorem} \label{Theorem Consistency}
    For an investor with initial wealth $x$ at time $t$, the optimal strategies and the maximum utilities under MV and MMV preferences are consistent if and only if $\Delta{L} \leq q^{mv}$ a.s..
\end{theorem}
\textbf{Proof.} When the solutions to MV and MMV problems are consistent, we have (1) $\pi^{mmv, \ast}(s) = \pi^{mv, \ast}(s)$, for any $s \in [t, T]$; (2) $C^{mmv} = C^{mv}$. Based on Lemma \ref{Lemma C and C' Relationship}, it is equivalent to $\Delta{L} \leq q^{mv}$ a.s.. Conversely, when the financial market admits $\Delta{L} \leq q^{mv}$ a.s., we have $q^{mmv} = q^{mv}$ and $C^{mmv} = C^{mv}$. Then, we have $\pi^{mmv,\ast}(s) = \frac{1}{q^{mv}} \left(e^{(s-t)r}x + \frac{e^{(T-t)C^{mv}-(T-s)r}}{\gamma} -X^{mmv,\ast}(s^-)\right)$, which has the same form as $\pi^{mv,\ast}(s)$ given by \eqref{MV Strategy}. Thus, $\pi^{mmv,\ast}(s) = \pi^{mv,\ast}(s)$ and $V_\gamma\left(X^{mmv,\ast}(T)\right) = U_\gamma\left(X^{mv,\ast}(T)\right)$. \hfill $\square$ 
\vskip 5pt

The economic interpretation of Theorem \ref{Theorem Consistency} is that when the financial market admits that $\Delta{L} \leq q^{mv}$ a.s., the investor under MV would not face the problem of non-monotonicity. However, when the condition $\Delta{L} \leq q^{mv}$ a.s. does not hold, the non-monotonicity of MV may induce the investor to behave irrationally. In such scenario, MMV can improve MV in portfolio selection and the investor under MMV satisfies the basic tenet of economic rationality. 

To delve deeper, MV and MMV investors behave differently when the jump size $\Delta{L}$ can be larger than $q^{mv}$. Both strategies entail a pre-determined target wealth. In the MMV scenario, the target is $e^{(s-t)r}x + \frac{e^{(T-t)C^{mmv}-(T-s)r}}{\gamma}$, whereas in the MV scenario, the target is $e^{(s-t)r}x + \frac{e^{(T-t)C^{mv}-(T-s)r}}{\gamma}$. Once the pre-determined target is exceeded, the MMV investor ceases further actions and holds a zero position in the risky asset. In contrast, the MV investor persists in pursuing the target, even if it entails maintaining a short position and incurring a negative profit, as elucidated in Propositions \ref{Proposition MV Stopping Time} and \ref{Proposition MMV Stopping Time}. This seemingly irrational strategy stems from the MV investor's perception of substantial upward profits as a form of risk, attributable to the variance component. Hence, despite both MV and MMV investors managing the jump risk and relinquishing investment opportunities, the MMV investor typically exhibits lower risk aversion. Moreover, $C^{mmv} > C^{mv}$ indicates that the pre-determined target of MMV exceeds that of MV. For visualized examples, refer to Section \ref{Section Numerical}.


Furthermore, we explore the connection between the wealth process and the monotone domain of MV preferences. As noted by \citet{strub2020note}, the terminal wealth under the optimal strategy ought to consistently remain within the monotone domain in any continuous semi-martingale market. Nevertheless, the case would be different when there are jumps in the market. 

\begin{proposition} \label{Proposition Monotone Domain}
    When the financial market admits that $\Delta{L} \leq q^{mv}$ a.s., both the terminal wealth under MV and MMV optimal strategies would keep within the monotone domain $\mathcal{G}_{\gamma}$ a.s.. Thus, MV and MMV optimal strategies are consistent. When the financial market violates that $\Delta{L} \leq q^{mv}$ a.s., both the terminal wealth under MV and MMV optimal strategies would fall outside the monotone domain $\mathcal{G}_{\gamma}$ with a positive probability. Thus, MV and MMV optimal strategies are different.
\end{proposition}
\textbf{Proof.} See Appendix \ref{Appendix Proof Monotone Domain} for the detailed proof. \hfill $\square$ 
\vskip 5pt

\begin{remark}
    The proposition gives us another strong reason to use MMV in portfolio selection. Due to the property of MV's domain of monotonicity (see Lemma 2.1 of \citet{maccheroni2009portfolio}), an FCFS can be extracted from $X^{mv,*}(T)$ when the condition $\Delta L\leq q^{mv}$ does not hold, which means that the MV investor can achieve a better outcome. Thus, taking MV in the jump-diffusion market is not economically reasonable. In the literature, \citet{cui2012better} first study the FCFS problem of MV in discrete and continuous-time financial markets. Then, \citet{bauerle2015complete} and \citet{strub2020note} conjecture that both completeness and continuity of asset prices are sufficient for the non-existence of FCFS. Meanwhile, MMV overcomes such problem directly by its monotonicity property. We cannot extract any FCFS from $X^{mmv,*}(T)$, because for any random variable $X \leq X^{mmv,*}(T)$ a.s., we have $V_{\gamma}(X) \leq V_{\gamma}(X^{mmv,*}(T))$. We conclude that the existence of $\Delta{L} > q^{mv}$ is a necessary condition for the existence of an FCFS in the jump-diffusion market. 
\end{remark}
\vskip 5pt


\subsection{Explanation Based on Pricing Operators}
\noindent
Based on the results in Subsection \ref{Subsection Consistency}, it is evident that the possible jump with size larger than $q^{mv}$ is the key factor that leads to the difference between MV and MMV. However, \citet{li2023optimal} show that discontinuity is not a sufficient condition for such inconsistency, although the jump size can also be extremely large within the insurance market. This prompts an inquiry into the underlying dissimilarities between the insurance and financial markets. Our exploration aims to uncover the fundamental reason for the inconsistency of MV and MMV, shedding light on the economic essence of MMV within this subsection.

\citet{vcerny2020semimartingale} compares the solutions to MV and MMV in general semi-martingale markets symbolically. He concludes that the consistency of MV and MMV is equivalent to that the variance-optimal martingale measure is not signed. Inspired by him, we study what is the exact measure $\mathbb{Q}^{\phi^{\ast}}$ under MMV preferences. 

\begin{proposition} \label{Proposition MMV Pricing Operator}
    $\mathbb{Q}^{\phi^{\ast}}$ is an absolutely continuous martingale measure and the corresponding pricing operator is given by
    \begin{equation}
        \mathrm{d}M^{mmv,\ast}(s) = M^{mmv,\ast}(s^-) \Bigg( -r \mathrm{d}s - \phi_1^{\ast}\mathrm{d}B(s) - \int_{-1}^{\infty} \phi_2^{\ast}(q)\widetilde{N}(\mathrm{d}s,\mathrm{d}q) \Bigg), \quad M^{mmv,\ast}(t) = 1. \label{MMV pricing operator} \notag
    \end{equation}
\end{proposition}
\textbf{Proof.} 
Under the probability measure $\mathbb{Q}^{\phi^{\ast}}$, we have 
\begin{align}
    \mathrm{d}S_1(s) & = S_1(s^-) \Bigg[\left(\mu + \lambda\xi_1\right) \mathrm{d}s + \sigma \left(\mathrm{d} B^{\phi^{\ast}}(s) - \phi_1^{\ast}\mathrm{d}s \right) + \int_{-1}^{\infty} q \left( \widetilde{N}^{\phi^{\ast}}(\mathrm{d}s,\mathrm{d}q) - \phi_2^{\ast}(q) \nu \left(\mathrm{d}q\right) \mathrm{d}s \right) \Bigg] \notag \\
    & = S_1(s^-) \Bigg( r \mathrm{d}s + \sigma \mathrm{d} B^{\phi^{\ast}}(s) + \int_{-1}^{\infty} q \widetilde{N}^{\phi^{\ast}}(\mathrm{d}s,\mathrm{d}q) \Bigg), \notag
    \end{align} 
which indicates that $\mathbb{Q}^{\phi^{\ast}}$ is actually an absolutely continuous martingale measure. Refer to \citet{delbaen1995existence} for details about absolutely continuous martingale measures. Besides, 
\begin{equation}
    \mathrm{d}M^{mmv,\ast}(s)S_1(s) = M^{mmv,\ast}(s^-)S_1(s^-) \Bigg[ \left(\sigma-\phi_1^*\right)\mathrm{d}B(s) + \int_{-1}^{\infty} \left(q-\phi_2^{\ast}(q)-q\phi_2^{\ast}(q)\right)\widetilde{N}(\mathrm{d}s,\mathrm{d}q) \Bigg],  \notag
\end{equation} which means that  $M^{mmv,\ast}$ is a pricing   operator.  $\phi_2^{\ast}(q) \leq 1$ ensures the non-negativity of $M^{mmv,\ast}$. \hfill $\square$ 
\vskip 5pt

\begin{proposition} \label{Proposition MV Pricing Operator}
    The pricing operator taken by the MV investor is given by
    \begin{equation}
        \mathrm{d}M^{mv,\ast}(s) = M^{mv,\ast}(s^-) \Bigg( -r \mathrm{d}s - \frac{\sigma}{q^{mv}}\mathrm{d}B(s) - \int_{-1}^{\infty} \frac{q}{q^{mv}} \widetilde{N}(\mathrm{d}s,\mathrm{d}q) \Bigg), \quad M^{mv,\ast}(t) = 1. \label{MV pricing operator} \notag
    \end{equation}
    When the financial market admits $\Delta{L} \leq q^{mv}$ a.s., $M^{mv,\ast}$ is the same as that given by Proposition \ref{Proposition MMV Pricing Operator}. But when $\Delta{L} \leq q^{mv}$ a.s. does not hold, $M^{mv,\ast}$ can be negative when $\Delta{L} > q^{mv}$. For such case, as it may assign a negative value for a positive return, it is not a reasonable pricing operator.
\end{proposition} 
\textbf{Proof.} Actually, the classical MV preference can be rewritten as
\begin{equation}
    U_{\gamma}(X) = \inf_{Y \in \mathcal{L}^2(\mathbb{P}), \mathbb{E}[Y]=1} \mathbb{E}^\mathbb{P} \left[ XY + \frac{1}{2\gamma} Y^2 \right] - \frac{1}{2\gamma}. \notag
\end{equation}
Based on such representation and Theorem \ref{Theorem MV Investor}, given $\pi^{mv,\ast}$ and $X^{mv,\ast}(T)$, the optimal $Y^{mv,\ast}$ under MV preferences is given by the Doléans-Dade exponential
\begin{equation}
    Y^{mv,\ast} = \mathcal{E}\Bigg(-\int_t^. \frac{\sigma}{q^{mv}} \mathrm{d}B(\tau) - \int_t^. \int_{-1}^{\infty} \frac{q}{q^{mv}} \widetilde{N} \left(\mathrm{d}\tau,\mathrm{d}q \right) \Bigg)_T. \notag
\end{equation}
When $\Delta{L} \leq q^{mv}$ holds, $Y^{mv,\ast} \geq 0$ a.s. and the probability measure $\mathbb{Q}^{mv, \ast}$ generated by $\frac{\mathrm{d}\mathbb{Q}^{mv, \ast}}{\mathrm{d}\mathbb{P}} =Y^{mv, \ast}$ is an absolutely continuous martingale measure, as under $\mathbb{Q}^{mv, \ast}$ we have 
\begin{equation}
    \mathrm{d}S_1(s) = S_1(s^-) \Bigg(r \mathrm{d}s + \sigma \mathrm{d} B^{mv, \ast}(s) + \int_{-1}^{\infty} q \widetilde{N}^{mv, \ast}(\mathrm{d}s,\mathrm{d}q) \Bigg),   \notag
\end{equation} 
where $\mathrm{d} B^{mv, \ast}(s) = \mathrm{d} B(s) + \frac{\sigma}{q^{mv}} \mathrm{d}s$ and $\widetilde{N}^{mv, \ast}(\mathrm{d}s,\mathrm{d}q) = \widetilde{N}(\mathrm{d}s,\mathrm{d}q) + \frac{q}{q^{mv}} \nu \left(\mathrm{d}q\right) \mathrm{d}s$. Thus, $M^{mv,\ast}$ is the corresponding pricing operator. \hfill $\square$ 
\vskip 5pt

\begin{remark}
    In an incomplete financial market, there are infinitely many absolutely continuous martingale measures. Given any possible absolutely continuous martingale measures $\mathbb{Q}$, we have
    \begin{equation}
        \mathbb{E}_{t,x}^{\mathbb{Q}} \left\{ X^{\pi}(T) + \frac{1}{2\gamma} \frac{\mathrm{d}\mathbb{Q}}{\mathrm{d}\mathbb{P}} \right\} - \frac{1}{2\gamma} = e^{(T-t)r}x + \frac{1}{2\gamma} \Bigg\{ \mathbb{E}_{t,x}^\mathbb{P} \Bigg[ \left(\frac{\mathrm{d}\mathbb{Q}}{\mathrm{d}\mathbb{P}}\right)^2 \Bigg] - 1 \Bigg\} = e^{(T-t)r}x + \frac{1}{2\gamma} \mathrm{Var}_{t,x}^{\mathbb{P}} \left[ \frac{\mathrm{d}\mathbb{Q}}{\mathrm{d}\mathbb{P}} \right], \notag
    \end{equation}
    for any admissible strategy $\pi$. As $\mathbb{Q}^{\phi^{\ast}}$ solves
    \begin{equation}
        \inf_{\mathbb{Q} \in \Delta^2(\mathbb{P})} \mathbb{E}_{t,x}^{\mathbb{Q}} \left\{ X^{\pi^{\ast}}(T) + \frac{1}{2\gamma} \frac{\mathrm{d}\mathbb{Q}}{\mathrm{d}\mathbb{P}} \right\} - \frac{1}{2\gamma}, \notag
    \end{equation}
   it has the minimum variance among all the square-integrable absolutely continuous martingale measures, verifying Theorem 2.5 of \citet{vcerny2020semimartingale}. Further, if we allow martingale measures to be signed \citep{vcerny2007structure}, $\mathbb{Q}^{mv, \ast}$ is actually the one with minimum variance among all the absolutely continuous signed martingale measures. Based on Proposition \ref{Proposition MV Pricing Operator}, when $\Delta{L} \leq q^{mv}$ a.s., $\mathbb{Q}^{mv, \ast}$ is not signed and thus consistent with $\mathbb{Q}^{\phi^{\ast}}$, verifying Theorem 5.4 of \citet{vcerny2020semimartingale}. In insurance markets, as jumps measuring claim payments are always non-positive \citep{schweizer1994approximating, schweizer1996approximation}, the variance-optimal martingale measure also keeps non-negative, showing again the consistency of MV and MMV. Besides, the stochastic process that is used to discount future cash flows of financial assets is usually called the pricing kernel (stochastic discount factor) and required to be strictly positive. Here, as $M^{mv,\ast}$ and $M^{mmv,\ast}$ can be non-positive, we follow \citet{dybvig1982mean} and call them pricing operators. 
\end{remark}
\vskip 5pt

From the perspective of the pricing operator, the MV investor optimizes his portfolio under the pricing operator given by Proposition \ref{Proposition MV Pricing Operator}, which would be negative when the jump is relatively large. Such pricing operator does not make sense for a rational investor. Such phenomenon is first discovered by \citet{dybvig1982mean}. By studying the CAPM in a single-period complete market, they find that the pricing operator can assign a negative value to a payoff which is strictly non-negative, resulting in an arbitrage opportunity. Actually, that is the fundamental reason for the non-monotonicity of MV. Based on Proposition \ref{Proposition MV Pricing Operator}, we find the same phenomenon in a continuous-time case when the financial market allows $\Delta{L} > q^{mv}$. 

Besides, the arbitrage opportunity in \citet{dybvig1982mean} requires the market portfolio generating an sufficiently large return, i.e., $\mathrm{Prob}(\widetilde{X}_m > \bar{X}_m + \frac{1}{\lambda_m}) > 0$, where $\lambda_m$ is the market price of risk. Similarly, we also find that the non-monotonicity of MV is induced by the jump larger than $q^{mv}$, where $q^{mv}$ is actually the inverse of the market price of risk in our model, i.e., $q^{mv} = \frac{1}{\lambda_m} = \frac{\sigma^2 + \lambda \xi_2^2 }{\mu - r + \lambda\xi_1}$. When the jump size $\Delta L$ can be larger than $q^{mv} = \frac{1}{\lambda_m}$, we have $\mathrm{Prob}(\frac{\Delta S_1}{S_1^-} > \mathbb{E}^\mathbb{P}\left[\frac{\Delta S_1}{S_1^-}\right] + \frac{1}{\lambda_m}) > 0$, where $\frac{\Delta S_1}{S_1^-}$ is the instantaneous payoff to the risky asset and $\mathbb{E}^\mathbb{P}\left[\frac{\Delta S_1}{S_1^-}\right] = \mathbb{E}^\mathbb{P}\left[\lambda \xi_1 \Delta t\right] = 0$. Thus, the result we get in a continuous-time incomplete model is completely comparable to the single-period complete model given by \citet{dybvig1982mean}, revealing the fundamental reason for MV's non-monotonicity. 

Further, \citet{dybvig1982mean} state that the arbitrage problem does not occur in the continuous-time log-normal diffusion model, as the market return cannot take on any large value over a short instant. Although they do not provide a proof, the result of \citet{strub2020note} just verifies the former's intuition that there is no FCFS  problem in any continuous market. Our result in the continuous-time discontinuous model just complements to them. Moreover, we find the channel of MMV to avoid the problem of non-monotonicity. Based on Proposition \ref{Proposition MMV Pricing Operator}, the MMV investor optimizes his portfolio under the pricing operator $M^{mmv, \ast}$, which is assigned to be zero when $\Delta{L} > q^{mmv}$. To conclude, MMV fixes MV by restricting the pricing operator to be non-negative, which is the essential reason for the inconsistency of MV and MMV. 

\section{Multi-Asset MMV Portfolio Selection} \label{Section Multi-Asset}
\noindent
In this section, we extend MMV portfolio selection theory from a single-asset market to a multi-asset market. For MMV investors, we verify the validity of the two-fund separation theorem, which is one of the most central results of the classical capital asset pricing model (CAPM) developed by \citet{sharpe1964capital}, \citet{lintner1965valuation} and \citet{mossin1966equilibrium}. Then, we establish the monotone capital asset pricing model (monotone CAPM) based on MMV, which is different from the classical CAPM based on MV. 

\subsection{Multi-Asset Market Model}\label{Section multi}
\noindent
We assume that the Brownian motions $\left\{B_j\right\}_{1\leq j\leq d}$ and the compound Poisson processes $\left\{L_i\right\}_{1\leq i\leq n}$ are mutually independent. The filtration $\left\{\mathcal{F}_t\right\}_{t\geq 0}$ is generated by these $d + n$ processes. Suppose that there are one risk-free asset and $n$ risky assets, with price processes given by
\begin{equation}
    \left\{
    \begin{aligned}
    & {\mathrm{d}S_0(t)} ={S_0(t)} r \mathrm{d} t, \\
    & \mathrm{d}S_i(t) ={S_i(t^-)}\Bigg( \mu_i \mathrm{d}t + \sum_{j = 1}^{d} \sigma_{ij} \mathrm{d} B_j(t) + \mathrm{d} \sum_{k = 1}^{n_i(t)} Q_{ik} \Bigg ), \quad i = 1, 2, \cdots, n.
    \end{aligned}
    \right. \label{Multi-Asset Financial Market}
\end{equation}
Assume  $\boldsymbol{\mu} = (\mu_1, \cdots, \mu_n)^T$, $\boldsymbol{\sigma}_i = (\sigma_{i1}, \cdots, \sigma_{id})^T$, $\boldsymbol{\sigma} = (\boldsymbol{\sigma}_1, \cdots, \boldsymbol{\sigma}_n)^T$. The L\'{e}vy process $L_i(t) = \sum \limits_{k = 1}^{n_i(t)} Q_{ik} = \int_0^t \int_{-1}^{\infty} q_i N_i(\mathrm{d}s,\mathrm{d}q_i)$ and $\lambda_i$ is the intensity associated with $n_i$. Besides, $\xi_{i1} = \mathbb{E}^{\mathbb{P}}[Q_{ik}]$, $\xi^2_{i2} = \mathbb{E}^{\mathbb{P}} [Q_{ik}^2]$ are finite moments of the jump sizes. At time $t$, the investor puts $\boldsymbol{\pi}(t) = (\pi_1(t), \cdots, \pi_n(t))^T$ amount into the risky assets and $X^{\pi}(t)-\sum_{i = 1}^{n}\pi_i(t)$ into the risk-free asset. 

For the sake of discussing MMV portfolio selection in a multi-asset financial market, here we also prove a generalized version of Theorem \ref{Theorem MMV V2 Representation}.
\begin{theorem} \label{Theorem MMV V3 Representation}
    Let $Y=\mathcal{E}(\boldsymbol{\phi})$ be the solution of SDE: 
	\begin{equation}
		\mathrm{d}Y^{\boldsymbol{\phi}}(t) = Y^{\boldsymbol{\phi}}(t^{-}) \left(-\sum_{j = 1}^{d}\phi_{1j}(t)\mathrm{d}B_j(t) -\sum_{i = 1}^{n} \int_{-\infty}^{\infty}\phi_{2i}(t,q)\widetilde{N}_i(\mathrm{d}t,\mathrm{d}q) \right), \quad Y^{\boldsymbol{\phi}}(0) = 1, \notag
	\end{equation}
    and the feasible set of $\boldsymbol{\phi} \triangleq \left(\phi_{11}, \cdots, \phi_{1d}, \phi_{21}, \cdots, \phi_{2n} \right)^T$ is
	\begin{equation}
		\boldsymbol{\varPhi} = \Big\{\boldsymbol{\phi}: \boldsymbol{\phi} \ \text{is predictable}, \ \phi_{2i}(t,q) \leq 1 \ \text{for all} \ \ (t,q) \in [0,T] \times \R,\ j=1,\cdots,n, \ Y^{\boldsymbol{\phi}}(T) = \mathcal{E}(\boldsymbol{\phi})_T \in \mathcal{M} \Big\}. \notag
	\end{equation} 
    Then, $V_\gamma(X)=\inf\limits_{Y \in \mathcal{M}_4}\mathbb{E}^\mathbb{P} \left[ XY + \frac{1}{2\gamma} Y^2\right] - \frac{1}{2\gamma}$, where $\mathcal{M}_4 = \Big\{ Y: Y \in \mathcal{M}, Y = \mathcal{E}(\boldsymbol{\phi})_T \ \text{for some} \ \boldsymbol{\phi} \in \boldsymbol{\varPhi} \Big\}$. 
\end{theorem}
\textbf{Proof.} See Appendix \ref{Appendix Proof MMV V3 Representation} for the detailed proof.\hfill $\square$ 
\vskip 5pt

Then, based on Theorem \ref{Theorem MMV V3 Representation}, MMV preferences can also be represented as \eqref{MMV Auxiliary Preference} in the multi-asset market. Just following the almost same procedure as Section \ref{Section MMV Portfolio Selection}, we establish the robust control problem, guess and verify the optimal control and value function. The only difference is that we need an assumption as follows.

\begin{assumption} \label{Assumption Q Existence}
    There exists a unique $\boldsymbol{q^{mmv}} = (q^{mmv}_1, \cdots, q^{mmv}_n)^T \in [-1, \infty)^n$ such that it solves
    \begin{equation}
        \textbf{Diag}\left(q^{mmv}_1, \cdots, q^{mmv}_n\right) \boldsymbol{\hat{\Omega}}^{-1}\left(\boldsymbol{\mu} - \boldsymbol{r} + \boldsymbol{\hat{\omega}} \right) = \boldsymbol{e}, \label{Multi Q Existence}
    \end{equation}
    where $\boldsymbol{\hat{\Omega}} = \boldsymbol{\sigma}\boldsymbol{\sigma}^T + \textbf{Diag}\left(\int_{-1}^{q^{mmv}_1}\widetilde{q}_1^2\nu_1(\mathrm{d}\widetilde{q}_1), \cdots, \int_{-1}^{q^{mmv}_n}\widetilde{q}_n^2\nu_n(\mathrm{d}\widetilde{q}_n)\right)$, $\boldsymbol{\hat{\omega}} = \left(\int_{-1}^{q^{mmv}_1}\widetilde{q}_1\nu_1(\mathrm{d}\widetilde{q}_1), \cdots, \int_{-1}^{q^{mmv}_n}\widetilde{q}_n\nu_n(\mathrm{d}\widetilde{q}_n)\right)^T$, and $\boldsymbol{e}$ is an n-dimensional column vector with all terms being 1. 
\end{assumption} 

\begin{remark}
    When processes of $S_i$ can be represented as 
    \begin{equation}
        \mathrm{d}S_i(t) ={S_i(t^-)}\Bigg( \mu_i \mathrm{d}t + \sigma_{i} \mathrm{d} B_i(t) + \mathrm{d} \sum_{k = 1}^{n_i(t)} Q_{ik} \Bigg ), \quad i = 1, 2, \cdots, n, \notag
    \end{equation}
    then $\boldsymbol{\sigma}$ is a diagonal matrix and $\boldsymbol{q^{mmv}}$ is the solution to Equation: 
    \begin{equation}
        \textbf{Diag}\left(q^{mmv}_1, \cdots, q^{mmv}_n\right) \textbf{Diag}\left(\sigma_1^{2} + \int_{-1}^{q^{mmv}_1}\widetilde{q}_1^2\nu_1(\mathrm{d}\widetilde{q}_1), \cdots, \sigma_n^{2} + \int_{-1}^{q^{mmv}_n}\widetilde{q}_n^2\nu_n(\mathrm{d}\widetilde{q}_n) \right)^{-1}\left(\boldsymbol{\mu} - \boldsymbol{r} + \boldsymbol{\hat{\omega}} \right) = \boldsymbol{e}, \notag
    \end{equation}
    which is equivalent to 
    \begin{equation}
        \frac{\sigma_i^{2} + \int_{-1}^{q^{mmv}_i}\widetilde{q}_i^2\nu_i(\mathrm{d}\widetilde{q}_i)}{\mu_i - r + \int_{-1}^{q^{mmv}_i}\widetilde{q}_i\nu_i(\mathrm{d}\widetilde{q}_i)} = q^{mmv}_i, \quad i = 1, 2, \cdots, n. \notag
    \end{equation}
    Based on Lemma \ref{Lemma Q Existence and Uniqueness}, we have the existence and uniqueness of $\boldsymbol{q^{mmv}}$. However, for the general case, it is a mathematical problem and we do not find out a way to show the same conclusion. Thus, we make the above assumption and we think the existence of $\boldsymbol{q^{mmv}}$ in general multi-asset markets is worthy of future study. 
\end{remark}
Now we are ready to state the main theorem solving the multi-dimensional MMV problem.

\begin{theorem} \label{Theorem Multi-Asset MMV investor}
    Suppose that Assumption \ref{Assumption Q Existence} is true across this section. Then, for an investor with initial wealth $x$ at time $t$, its optimal investment strategy under MMV preferences in the multi-asset market is 
    \begin{align}
        \boldsymbol{\pi}_{MA}^{mmv,\ast}(s) = \boldsymbol{\hat{\Omega}}^{-1}\left(\boldsymbol{\mu} - \boldsymbol{r} + \boldsymbol{\hat{\omega}} \right) \max\left\{e^{(s-t)r}x + \frac{e^{(T-t)C^{mmv}_{MA}-(T-s)r}}{\gamma} -X_{MA}^{mmv,\ast}(s^-), 0 \right\}, \label{Multi-Asset MMV Strategy}
    \end{align}
     for any $s \in [t, T]$, where $\boldsymbol{q^{mmv}}$ is the solution of Equation \eqref{Multi Q Existence}, $C_{MA}^{mmv} = \left(\boldsymbol{\mu} - \boldsymbol{r} + \boldsymbol{\hat{\omega}} \right)^T \boldsymbol{\hat{\Omega}}^{-1} \left(\boldsymbol{\mu} - \boldsymbol{r} + \boldsymbol{\hat{\omega}} \right) + \sum_{i = 1}^{n} \int_{q_i^{mmv+}}^{\infty} \nu_i(\mathrm{d}\widetilde{q}_i)$, and $X_{MA}^{mmv,\ast}$ denotes the wealth process under the strategy $\boldsymbol{\pi}_{MA}^{mmv,\ast}$. 
\end{theorem}

The proof keeps almost the same as that of Theorem \ref{Theorem MMV investor}. When $n = 1$, Theorem \ref{Theorem Multi-Asset MMV investor} degenerates to Theorem \ref{Theorem MMV investor}. Similar to Proposition \ref{Proposition MMV Expression}, we write down the explicit form of the wealth process.  

\begin{proposition} \label{Proposition Multi-Asset MMV Expression}
    Define a stopping time $T_{\boldsymbol{q^{mmv}}} \triangleq \inf \Big\{s > t: \Delta L_i(s) \geq q^{mmv}_i \ \text{for some} \ i \Big\}$. When $t \leq s \leq T$, the explicit form of $X_{MA}^{mmv,\ast}(s)$ can be given by
    \begin{align}
        X^{mmv,\ast}(s) & = e^{(s-t)r}x + \frac{1}{\gamma}e^{(T-t)C^{mmv}_{MA}-(T-s)r} - 2e^{(T-s)(C^{mmv}_{MA}-r)}Y_{MA}^{\boldsymbol{\phi}^{\ast}}(s) \one_{\{t \leq s < T_{\boldsymbol{q^{mmv}}}\}} \notag \\
        & + 2e^{(T-T_{\boldsymbol{q^{mmv}}})C^{mmv}_{MA}- (T-s)r} \sum_{i = 1}^{n} \left(\frac{\Delta L_i(T_{\boldsymbol{q^{mmv}}})}{q^{mmv}_i} - 1 \right) Y_{MA}^{\boldsymbol{\phi}^{\ast}}(T_{\boldsymbol{q^{mmv}}}^-) \one_{\{s \geq T_{\boldsymbol{q^{mmv}}}\}}, \notag
    \end{align}
    where $Y_{MA}^{\boldsymbol{\phi}^{\ast}}(s)$ can be expressed as the form of the Doléans-Dade exponential
    \begin{equation}
    Y_{MA}^{\boldsymbol{\phi}^{\ast}}(s) = \frac{1}{2\gamma} \mathcal{E}\Bigg(- \sum_{j = 1}^{d} \int_t^. \phi_{1j}^{\ast} \mathrm{d}B_j(\tau) - \sum_{i = 1}^{n} \int_t^. \int_{-1}^{\infty} \phi_{2i}^{\ast}(q_i) \widetilde{N}_i \left(\mathrm{d}\tau,\mathrm{d}q_i \right) \Bigg)_s, \notag
    \end{equation}
    $\boldsymbol{\phi}^{\ast} = \left(\boldsymbol{\phi}_{1}^{\ast T}, \boldsymbol{\phi}_{2}^{\ast T}(q)\right)^T$ is given by
    \begin{equation}
        \left\{
        \begin{aligned}
        & \boldsymbol{\phi}_{1}^{\ast} = \left(\phi_{11}^{\ast},
        \cdots, \phi_{1d}^{\ast} \right)^T = \boldsymbol{\sigma}^T  \boldsymbol{\hat{\Omega}}^{-1}\left(\boldsymbol{\mu} - \boldsymbol{r} + \boldsymbol{\hat{\omega}} \right), \\
        & \boldsymbol{\phi}_{2}^{\ast}(q) = \left(\phi_{21}^{\ast}(q), \cdots, \phi_{2n}^{\ast}(q) \right)^T, \quad \phi_{2i}^{\ast}(q) = \min \Big\{ \boldsymbol{e}_i^T \boldsymbol{\hat{\Omega}}^{-1}\left(\boldsymbol{\mu} - \boldsymbol{r} + \boldsymbol{\hat{\omega}}\right) q, 1 \Big\}, \quad -1 \leq q < \infty, 
        \end{aligned}
        \right. \notag
    \end{equation}
    and $\boldsymbol{e}_i$ is an n-dimensional column vector with the i-th term being 1 and other terms being 0. 
\end{proposition}

Further, we establish the relationship between MV and MMV in the multi-asset financial market. We omit the proof here as it keeps almost the same as that of Theorem \ref{Theorem Consistency}. 

\begin{theorem}
    In the multi-asset financial market, the optimal strategies and the maximum utilities under MV and MMV preferences are consistent if and only if the market admits that $\Delta\boldsymbol{L} \leq \boldsymbol{q^{mv}}$ a.s., where $\Delta\boldsymbol{L} = \left(\Delta L_1, \cdots, \Delta L_n\right)^T$, and $\boldsymbol{q^{mv}} = (q^{mv}_1, \cdots, q^{mv}_n)^T$ is the solution to Equation: $\textbf{Diag}\left(q^{mv}_1, \cdots, q^{mv}_n\right) \boldsymbol{\bar{\Omega}}^{-1} \left(\boldsymbol{\mu} - \boldsymbol{r} + \boldsymbol{\bar{\omega}} \right)= \boldsymbol{e}$, with $ \boldsymbol{\bar{\Omega}} = \boldsymbol{\sigma}\boldsymbol{\sigma}^T + \textbf{Diag}\left(\lambda_1\xi_{12}^2, \cdots, \lambda_n \xi_{n2}^2\right)$, $\boldsymbol{\bar{\omega}} = \left(\lambda_1\xi_{11}, \cdots, \lambda_n\xi_{n1}\right)^T$. 
\end{theorem}

We give the pricing operator under MMV in the multi-asset market, which is comparable with that in Proposition \ref{Proposition MMV Pricing Operator}. The pricing operator keeps non-negative, making the preference monotone. Also, we omit the proof as it keeps the same as that of Proposition \ref{Proposition MMV Pricing Operator}. 

\begin{proposition} 
    $\mathbb{Q}_{MA}^{\boldsymbol{\phi}^{\ast}}$ defined by $\frac{\mathrm{d}\mathbb{Q}_{MA}^{\boldsymbol{\phi}^{\ast}}}{\mathrm{d}\mathbb{P}} = 2\gamma Y_{MA}^{\boldsymbol{\phi}^{\ast}}(T)$ is an absolute continuous martingale measure and the corresponding pricing operator is given by
    \begin{equation}
        \mathrm{d}M_{MA}^{mmv,\ast}(s) = M_{MA}^{mmv,\ast}(s^-) \Bigg( -r \mathrm{d}s - \sum_{j = 1}^{d} \phi_{1j}^{\ast} \mathrm{d}B_j(s) - \sum_{i = 1}^{n} \int_{-1}^{\infty} \phi_{2i}^{\ast}(q_i) \widetilde{N}_i \left(\mathrm{d}s,\mathrm{d}q_i \right) \Bigg), \quad M_{MA}^{mmv,\ast}(t) = 1. \label{Multi-Asset MMV pricing operator} \notag
    \end{equation}
\end{proposition}

\subsection{Monotone Capital Asset Pricing Model}
\noindent
So far, we have solved the MMV portfolio selection in the multi-asset financial market. Next, we follow the classical mean-variance analysis and show the asset pricing implication in the MMV setting. We begin by establishing the two-fund separation theorem. 

\begin{theorem} \label{Theorem MMV Two-Fund Separation}
    Given $\gamma, \gamma^{\prime} > 0$. If $\boldsymbol{\pi}_{MA}^{mmv,\ast}$ given in Theorem \ref{Theorem Multi-Asset MMV investor} solves the MMV portfolio selection problem for an investor with uncertainty aversion coefficient $\gamma$, then $\frac{\gamma}{\gamma^{\prime}}\boldsymbol{\pi}_{MA}^{mmv,\ast}$ solves the problem for an investor with uncertainty aversion coefficient $\gamma^{\prime}$. Thus, there exists a market portfolio, with proportions of risky assets given by
    \begin{equation}
        \boldsymbol{\Pi}^{mkt} = \frac{1}{\boldsymbol{e}^T \boldsymbol{\pi}_{MA}^{mmv,\ast}}\boldsymbol{\pi}_{MA}^{mmv,\ast} = \frac{1}{\boldsymbol{e}^T \boldsymbol{\hat{\Omega}}^{-1}\left(\boldsymbol{\mu} - \boldsymbol{r} + \boldsymbol{\hat{\omega}} \right)} \boldsymbol{\hat{\Omega}}^{-1}\left(\boldsymbol{\mu} - \boldsymbol{r} + \boldsymbol{\hat{\omega}} \right),
    \end{equation}
    independent of time. Under MMV preferences, all investors hold a combination of the risk-free asset and the market portfolio, i.e., the two-fund separation theorem holds. 
\end{theorem}
\vskip 5pt

The proof of Theorem \ref{Theorem MMV Two-Fund Separation} is obvious based on the expression \eqref{Multi-Asset MMV Strategy}. However, its economic implication is important. Theoretically, it indicates that all MMV investors should hold a portfolio of risky assets proportional to the market portfolio. Empirically, parameters in the continuous-time case can be estimated more precisely. The equilibrium composition of the market portfolio $\boldsymbol{\Pi}^{mkt}$ can be better determined compared to the discrete case (see Section 5 of \citet{maccheroni2009portfolio}). 

Based on the two-fund separation theorem, we study the asset pricing theory based on the market portfolio. According to the classical CAPM based on MV, the risk premium of each asset has a linear relationship with the risk premium of the market portfolio. \citet{maccheroni2009portfolio} study the asset pricing based on MMV and propose a monotone CAPM in a single-period model. Here, we extend their result to the continuous-time model and get an explicit expression of the beta. Consider the wealth process of the market portfolio $X^{mkt} \triangleq \Big\{X^{mkt}(s);t \leq s \leq T \Big\}$ defined as
\begin{equation}
    \mathrm{d} X^{mkt}(s) = \sum_{i = 1}^{n} X^{mkt}(s^-) \Pi_i^{mkt}\Bigg( \mu_i \mathrm{d}s + \sum_{j = 1}^{d} \sigma_{ij} \mathrm{d} B_j(s) + \int_{-1}^{\infty} q_i N_i(\mathrm{d}s, \mathrm{d}q_i) \Bigg ). \notag
\end{equation} 
Then, we have the following capital asset pricing theorem.
\begin{theorem} \label{Theorem MMV CAPM}
    Denote $R_i^{\tau} = \frac{S_i(T)}{S_i(t)}$ as the return of the risky asset $i$ during the period $\tau = T - t$, $R_{mkt}^{\tau} = \frac{X^{mkt}(T)}{X^{mkt}(t)}$ as the return of the market portfolio and $R_f^{\tau} = e^{(T-t)r}$ as the risk-free return. Then, we have the monotone capital asset pricing model (monotone CAPM)
    \begin{equation}
        \mathbb{E}_{t}^{\mathbb{P}}\big[R_{i}^{\tau}\big] - R_f^{\tau} = \beta_i^{mmv} \Big( \mathbb{E}_{t}^{\mathbb{P}}\big[R_{mkt}^{\tau}\big] - R_f^{\tau}\Big), \label{MMV CAPM}
    \end{equation}
    where the coefficient is given by
    \begin{equation}
        \beta_i^{mmv} = \frac{\mathrm{Cov}_{t}^{\mathbb{P}}\Big(Y_{MA}^{\phi^{\ast}}(T), R_{i}^{\tau}\Big)}{\mathrm{Cov}_{t}^{\mathbb{P}}\Big(Y_{MA}^{\phi^{\ast}}(T), R_{mkt}^{\tau}\Big)} = \frac{\mathrm{Cov}_{t}^{\mathbb{P}}\Big(M_{MA}^{mmv,\ast}(T), R_{i}^{\tau}\Big)}{\mathrm{Cov}_{t}^{\mathbb{P}}\Big(M_{MA}^{mmv,\ast}(T), R_{mkt}^{\tau}\Big)}. \label{MMV Beta}
    \end{equation}
\end{theorem}
\textbf{Proof.} For each single asset $i$, we have
\begin{align}
    \mathrm{Cov}_{t}^{\mathbb{P}}\Big(Y_{MA}^{\phi^{\ast}}(T), R_{i}^{\tau}\Big) & = \mathbb{E}_{t}^{\mathbb{P}}\big[Y_{MA}^{\phi^{\ast}}(T) R_{i}^{\tau}\big] - \mathbb{E}_{t}^{\mathbb{P}}\big[Y_{MA}^{\phi^{\ast}}(T)\big] \mathbb{E}_{t}\big[R_{i}^{\tau}\big] = \frac{1}{2\gamma} \mathbb{E}_{t}^{\mathbb{Q}_{MA}^{\phi^{\ast}}}\big[R_{i}^{\tau}\big] - \frac{1}{2\gamma}\mathbb{E}_{t}^{\mathbb{P}}\big[R_{i}^{\tau}\big] \notag \\
    & = \frac{1}{2\gamma} R_f^{\tau} - \frac{1}{2\gamma}\mathbb{E}_{t}^{\mathbb{P}}\big[R_{i}^{\tau}\big] < 0. \notag
\end{align} 
As the market portfolio is consisted of all single stocks, we also have
\begin{equation}
    \mathrm{Cov}_{t}^{\mathbb{P}}\Big(Y_{MA}^{\phi^{\ast}}(T), R_{mkt}^{\tau}\Big) = \frac{1}{2\gamma} R_f^{\tau} - \frac{1}{2\gamma}\mathbb{E}_{t}^{\mathbb{P}}\big[R_{mkt}^{\tau}\big] < 0. \notag
\end{equation}
Then
\begin{equation}
    \frac{\mathbb{E}_{t}^{\mathbb{P}}\big[R_{i}^{\tau}\big] - R_f^{\tau}}{\mathbb{E}_{t}^{\mathbb{P}}\big[R_{mkt}^{\tau}\big] - R_f^{\tau}} = \frac{\mathrm{Cov}_{t}^{\mathbb{P}}\Big(Y_{MA}^{\phi^{\ast}}(T), R_{i}^{\tau}\Big)}{\mathrm{Cov}_{t}^{\mathbb{P}}\Big(Y_{MA}^{\phi^{\ast}}(T), R_{mkt}^{\tau}\Big)} =\frac{\mathrm{Cov}_{t}^{\mathbb{P}}\Big(M_{MA}^{mmv,\ast}(T), R_{i}^{\tau}\Big)}{\mathrm{Cov}_{t}^{\mathbb{P}}\Big(M_{MA}^{mmv,\ast}(T), R_{mkt}^{\tau}\Big)}, \notag
    \end{equation}
thus the relationship \eqref{MMV CAPM} holds. \hfill $\square$ 

Further, terms in \eqref{MMV CAPM} and \eqref{MMV Beta} used for asset pricing can be explicitly expressed as follows. 

\begin{proposition}
    For the monotone CAPM, $R_{mkt}^{\tau}$ can be explicitly expressed as
    \begin{align}
        R_{mkt}^{\tau} & = e^{\frac{\left(\boldsymbol{\mu} - \boldsymbol{r} + \boldsymbol{\hat{\omega}} \right)^T \boldsymbol{\hat{\Omega}}^{-1}\left(\boldsymbol{\mu} + \boldsymbol{\hat{\omega}} \right)}{\boldsymbol{1}^T \boldsymbol{\hat{\Omega}}^{-1}\left(\boldsymbol{\mu} - \boldsymbol{r} + \boldsymbol{\hat{\omega}} \right)}(T-t)} \mathcal{E} \Bigg( \frac{1}{\boldsymbol{1}^T \boldsymbol{\hat{\Omega}}^{-1}\left(\boldsymbol{\mu} - \boldsymbol{r} + \boldsymbol{\hat{\omega}} \right)} \Bigg( \sum_{j = 1}^{d} \int_t^. \phi_{1j}^{\ast} \mathrm{d}B_j(s) \notag \\
        & + \sum_{i = 1}^{n} \int_t^. \int_{-1}^{\infty} \phi_{2i}^{\ast}(q_i) \widetilde{N}_i \left(\mathrm{d}s,\mathrm{d}q_i \right) + \sum_{i = 1}^{n} \int_t^. \int_{q_i^{mmv+}}^\infty \Big( \boldsymbol{e}_i^T \boldsymbol{\hat{\Omega}}^{-1}\left(\boldsymbol{\mu} - \boldsymbol{r} + \boldsymbol{\hat{\omega}}\right) q_i - 1\Big) \widetilde{N}_i \left(\mathrm{d}s,\mathrm{d}q_i \right)  \Bigg) \Bigg)_T, \notag
    \end{align}
    and $M_{MA}^{mmv,\ast}(T)$ can be explicitly expressed as 
    \begin{equation}
        M_{MA}^{mmv,\ast}(T) = e^{-(T-t)r} \mathcal{E}\Bigg(- \sum_{j = 1}^{d} \int_t^. \phi_{1j}^{\ast} \mathrm{d}B_j(s) - \sum_{i = 1}^{n} \int_t^. \int_{-1}^{\infty} \phi_{2i}^{\ast}(q_i) \widetilde{N}_i \left(\mathrm{d}s,\mathrm{d}q_i \right)   \Bigg)_T. \notag
    \end{equation}
\end{proposition}

When the condition $\Delta\boldsymbol{L} \leq \boldsymbol{q^{mv}}$ a.s. does not hold, the investor takes different pricing operators under MV and MMV, resulting in different portfolio selection choices. As shown in \eqref{MMV Beta}, both the pricing operator and the market portfolio make the beta ($\beta_{i}^{mmv}$) different from the one based on MV. It is also comparable to the result based on the single-period model. As \citet{maccheroni2009portfolio} say, the monotone CAPM is arbitrage free and the beta can be inferred from the market data. Here, we solve out the beta more explicitly and relevant market parameters can be estimated more precisely in the continuous-time model. Such a result can be a theoretical basis for future empirical tests of the effectiveness of the monotone CAPM. 

\begin{remark} 
    As shown in \eqref{MMV CAPM}, the risk premium for each asset still maintains a linear relationship with the risk premium for the market portfolio, consistent with the traditional view that the idiosyncratic risk can be fully diversified and only the market risk should be priced. The primary disparities between classical CAPM and monotone CAPM lie in the altered beta and market portfolio. When $\Delta\boldsymbol{L} \leq \boldsymbol{q^{mv}}$ a.s., the pricing operator $M_{MA}^{mmv,\ast}$ remains identical to $M_{MA}^{mv,\ast}$, thereby preserving the beta and market portfolio. However, in cases where $\Delta\boldsymbol{L} \leq \boldsymbol{q^{mv}}$ a.s. does not hold, $M_{MA}^{mv,\ast}(s)$ turns negative when $\Delta L_i(s) > q^{mv}_i$ for some $i$ and $s$. Instead, $M_{MA}^{mmv,\ast}(s)$ becomes zero when $\Delta L_i(s) > q^{mmv}_i$ for some $i$ and $s$. Even though $\Delta L_i(s) > q^{mmv}_i$ does not occur for any $i$ or $s$ in a sample path, $M_{MA}^{mmv,\ast}(T)$ diverges from $M_{MA}^{mv,\ast}(T)$. Consequently, both beta and market portfolio undergo alterations.
\end{remark}
\vskip 5pt

\section{MMV in Constrained Trading Model}{\label{Section Constraint}}
\noindent
Due to the two-fund separation given by Theorem \ref{Theorem MMV Two-Fund Separation}, it is reasonable for us to regard the risky asset in our original model as the aggregate market portfolio for simplicity. Recently, there are works studying MMV portfolio selection problems when the trading strategy is constrained \citep{shen2022cone, hu2023constrained}. Following a similar method used in \cite{strub2020note}, \citet{du2023monotone} generalize the constrained trading market model and conclude that the solutions to MV and MMV problems are still consistent whenever asset prices are continuous. However, there is no work studying the case that the market is discontinuous and the trading strategy is constrained. The relationship between MV and MMV in such scenario is still unclear. 

Based on our original jump-diffusion model, we solve MV and MMV problems with no-short constraint. That means that we need the admissible trading strategy to satisfy $\pi \geq 0$. Under MV preferences, it is a classical problem in the literature and we have to study the viscosity solution of the HJB equation \citep{li2002dynamic}. Here, we omit the detailed process and directly give the result. 

\begin{theorem} \label{Theorem MV Trading Constraint}
    For an investor with initial wealth $x$ at time $t$ in the market with no-short constraint, its optimal investment strategy under MV preferences is
    \begin{equation}
        \pi_{NS}^{mv,\ast}(s) = \left\{
        \begin{aligned}
            & \frac{1}{q^{mv}}\left(e^{(s-t)r}x + \frac{e^{(T-t)C^{mv}-(T-s)r}}{\gamma}-X_{NS}^{mv,\ast}(s^-)\right), & \text{if} \quad t \leq s < T_{\pi^{\ast}} \wedge T, \\
            & 0, & \text{if} \quad T_{\pi^{\ast}} \wedge T \leq s \leq T,
        \end{aligned}
        \right.  \notag
    \end{equation}
    where $T_{\pi^{\ast}} \triangleq \inf \Big\{s > t: X_{NS}^{mv,\ast}(s) \geq e^{(s-t)r}x + \frac{e^{(T-t)C^{mv}-(T-s)r}}{\gamma} \Big\}$. 
\end{theorem}

\begin{proposition}
    Actually, we have $T_{\pi^{\ast}} = T_{q^{mv}}$. Under MV preferences, the maximum utility and efficient frontier in the market with no-short constraint are the same as those in the market without no-short constraint. But the optimal strategies in the two models are consistent if and only if $T_{\pi^{\ast}} \geq T$. When the financial market admits that $\Delta{L} \leq q^{mv}$ a.s., the investor would not meet the trading constraint and the optimal strategies in the two models are consistent. 
\end{proposition}

For the MMV problem, based on the results in Theorem \ref{Theorem MMV investor} and Proposition \ref{Proposition MMV Stopping Time}, we know that the optimal strategy in the market without no-short constraint has already satisfied the requirement of non-negativity. Thus, it must be the optimal strategy in the market with the constraint. 

\begin{theorem} \label{Theorem MMV Trading Constraint}
    For an investor with initial wealth $x$ at time $t$ in the market with no-short constraint, its optimal investment strategy under MMV preferences is 
    \begin{align}
        \pi_{NS}^{mmv,\ast}(s) = \frac{1}{q^{mmv}} \left(e^{(s-t)r}x + \frac{e^{(T-t)C^{mmv}-(T-s)r}}{\gamma} - X_{NS}^{mmv,\ast}(s^-)\right), \quad t \leq s \leq T, \notag
    \end{align}
     which is the same as that in the market without no-short constraint given in Theorem \ref{Theorem MMV investor}. 
\end{theorem}

\begin{proposition}
    With no-short constraint, the optimal strategies and the maximum utilities under MV and MMV preferences are consistent if and only if $\Delta{L} \leq q^{mv}$ a.s.. 
\end{proposition}

\begin{remark}
    The similarity between the strategies in Theorem \ref{Theorem MV Trading Constraint} and Theorem \ref{Theorem MMV Trading Constraint} (Theorem \ref{Theorem MMV investor}) is that, the investor changes to keep zero amount of the risky asset after the relatively large jump occurs and then the wealth grows at the risk-free rate till the final stage. However, there are still differences because the investor takes different pricing operators under two preferences when the market violates $\Delta{L} \leq q^{mv}$ a.s., which also leads to the difference in the jump size ($q^{mv}$ or $q^{mmv}$) that prompts them to stop the risky investment. 
\end{remark}
\vskip 5pt
\section{Numerical Analysis}{\label{Section Numerical}}
\noindent
In this section, we provide several examples to compare the solutions to MV and MMV problems numerically. We also consider the MV strategy with no-short constraint given by Theorem \ref{Theorem MV Trading Constraint} and compare it with the MMV strategy. Based on Theorem \ref{Theorem Consistency}, when the financial market violates $\Delta L \leq q^{mv}$ a.s., MMV improves MV. In the following discussion, we set the basic parameters as: $x_0 = 1$, $t = 0$, $T = 1$, $\gamma = 2$, $\mu = 0.25$, $r = 0.05$, $\sigma = 0.15$. Parameters of the jump can be different in each scenario. 

\subsection{Constantly Distributed Jump}
\noindent
Suppose that the size of the return jump $Q_i$ is actually a constant, i.e., $Q_i = Q_0$. Then, $\xi_1 = Q_0$, $\xi_2^2 = Q_0^2$ and $q^{mv} = \frac{\sigma^2 + \lambda Q_0^2}{\mu - r + \lambda Q_0}$. By simple calculation, we have that $q^{mmv} = \frac{\sigma^2 + \lambda Q_0^2}{\mu - r + \lambda Q_0}$, if $Q_0 \leq \frac{\sigma^2}{\mu - r}$;  $q^{mmv} = \frac{\sigma^2}{\mu - r}$, if $Q_0 > \frac{\sigma^2}{\mu - r}$. We are interested in the scenario when two solutions are different, i.e., $Q_0 > \frac{\sigma^2}{\mu - r}$. Given $\lambda = 2$, $Q_0 = \xi_1 = \xi_2 = 0.2 > \frac{\sigma^2}{\mu-r}$, we have $q^{mv} = 0.1708$, $q^{mmv} = 0.1125$, $C = \frac{(\mu-r+\lambda Q_0)^2}{\sigma^2 + \lambda Q_0^2} = 3.5122$ and $C^{mmv} = \frac{(\mu-r)^2}{\sigma^2} + \lambda = 3.7778$. 

Curves in Figure \ref{Constant Distributed} simulate the dynamic wealth processes and the corresponding investment strategies, following the same sample path. Also, the estimated density distribution (by 1000 experiments) of investors' terminal wealth is presented in the third sub-figure. The blue solid curve corresponds to MV, the green one corresponds to MV with no-short constraint, and the red one corresponds to MMV. They may coincide with each other and we will make explanations. Besides, the blue (red) dashed curve represents the pre-determined target o that drives the investor to short sell (hold zero position) of the risky asset under MV (MMV). 

Several facts can be observed. First, before the jump occurs, three investment strategies keep positive and wealth processes do not leave the pre-determined upper bound of wealth (the blue or red dashed curve). As no-short constraint is not violated, the wealth process (investment strategy) under MV just coincides with the process (strategy) under MV with no-short constraint. Second, when the jump occurs, three wealth processes all jump across the bound. While the investment strategy under MV becomes negative and the wealth process continues to fluctuate, both strategies under MV with no-short constraint and MMV become zero and both wealth processes just grow at the risk-free rate till the final stage. Further, it seems that most of the time the investment in the risky asset under MMV is greater than that under MV. And the wealth process under MMV seems to be more volatile, indicating that MMV is less uncertainty averse than MV. From the distribution of end-of-period wealth, the wealth distribution under MMV appears to have a higher mean, which is attributed to the monotonicity property. Also, the distribution under MMV has a thicker tail, which also confirms that MMV investor exhibits lower uncertainty aversion. 


\subsection{Uniformly Distributed Jump}
\noindent
Suppose that the size of the return jump $Q_i$ follows a uniform distribution, i.e., $Q_i \sim U[Q_d, Q_u]$, where $Q_d \geq -1$. Assume $\Delta Q = Q_u - Q_d$. Then, $\xi_1 = \frac{Q_d + Q_u}{2}$, $\xi_2^2 = \frac{Q_d^2 + Q_d Q_u + Q_u^2}{3}$. The unique solution is $q^{mmv} = \frac{\sigma^2 + \lambda \xi_2^2}{\mu - r + \lambda \xi_1}$, if $Q_u \leq \frac{\sigma^2 + \lambda \xi_2^2}{\mu - r + \lambda \xi_1}$; $q^{mmv} = \frac{\sigma^2}{\mu - r}$, if $Q_d \geq \frac{\sigma^2}{\mu - r}$. When $Q_u > \frac{\sigma^2 + \lambda \xi_2^2}{\mu - r + \lambda \xi_1}$ and $Q_d < \frac{\sigma^2}{\mu - r}$, the zero point $q^{mmv}$ is given by solving the equation
\begin{equation}
    \frac{\lambda}{6}q^{mmv3} + \left[\left(\mu-r\right)\Delta Q - \frac{\lambda}{2}Q_d^2\right]q^{mmv} + \frac{\lambda}{3}Q_d^3 - \sigma^2 \Delta Q = 0. \notag
\end{equation}
We are interested in the scenario when the two solutions are different, i.e., $Q_u > \frac{\sigma^2 + \lambda \xi_2^2}{\mu - r + \lambda \xi_1}$. Given $\lambda = 2$, $Q_d = -0.1, Q_u = 0.5$, we have $\xi_1 = 0.2$, $\xi_2 = 0.2646$, $q^{mv} = 0.2708$ and $q^{mmv} = 0.1232$. Further, $C = \frac{\mu - r + \frac{\lambda}{2}(Q_u + Q_d)}{q^{mv}} = 2.2154$ and $C^{mmv} = \frac{\mu-r+\frac{\lambda}{2}(Q_u + Q_d) - \frac{\lambda}{2}\frac{\left(Q_u - q^{mmv}\right)^2}{\Delta Q}}{q^{mmv}} = 2.9499$. 

Curves in Figure \ref{Uniform Distributed} simulate the dynamic wealth processes, the corresponding investment strategies and the wealth distribution. We observe the similar pattern as in Figure \ref{Constant Distributed} but with little differences. There are two notable jumps in this sample path. The size of the first jump is larger than $q^{mmv}$ but smaller than $q^{mv}$, resulting that the MMV investment strategy becomes zero but the MV strategy and that with no-short constraint are still positive. However, the size of the second jump is larger than $q^{mv}$, making the wealth process under MV exceeding the upper bound and the investment strategy being negative. Correspondingly, the strategy under MV with no-short constraint becomes zero and the wealth keep growing at the risk-free rate. Consistently, the wealth under MMV is more likely to reach larger extreme values. 


\subsection{Exponentially Distributed Jump}
\noindent
Suppose that the size of the return jump $Q_i$ follows an exponential distribution, i.e., $Q_i \sim Exp(\theta ; Q_d)$, where $Q_d \geq -1$. $g(q) = 0$ when $q < Q_d$; $g(q) = \theta e^{-\theta (q - Q_d)}$ when $q \geq Q_d$. Then, $\xi_1 = \frac{1}{\theta} + Q_d$, $\xi_2^2 = \left(\frac{1}{\theta} + Q_d\right)^2 + \frac{1}{\theta^2}$. The unique solution $q^{mmv} = \frac{\sigma^2}{\mu - r}$, if $Q_d \geq \frac{\sigma^2}{\mu - r}$. When $Q_d < \frac{\sigma^2}{\mu-r}$, the zero point $q^{mmv}$ is given by solving 
\begin{equation}
    \sigma^2 - (\mu - r)q^{mmv} + \lambda \left(Q_d^2 + \frac{2}{\theta}Q_d + \frac{2}{\theta^2}\right) - \lambda \left(Q_d + \frac{1}{\theta}\right)q^{mmv} + \lambda \left(\frac{1}{\theta}q^{mmv} + \frac{2}{\theta^2}\right)e^{-\theta(q^{mmv}-Q_d)} = 0. \notag
\end{equation}
Given $Q_d = -0.3, \theta = 2$, we have $\xi_1 = 0.2, \xi_2 = 0.5385$, $q^{mv} = 1.0042$, $q^{mmv} = 0.1324$. Further, $C = \frac{\mu - r + \lambda \left(\frac{1}{\theta} + Q_d\right)}{q^{mv}} = 0.5975$ and $C^{mmv} = \frac{\mu-r+ \lambda \left(\frac{1}{\theta} + q^{mmv} - \frac{1}{\theta}e^{-\theta(q^{mmv}-Q_d)} \right)}{q^{mmv}} = 1.0435$. 

\begin{figure}[htbp]
    \FIGURE
    {\includegraphics[width=1\linewidth]{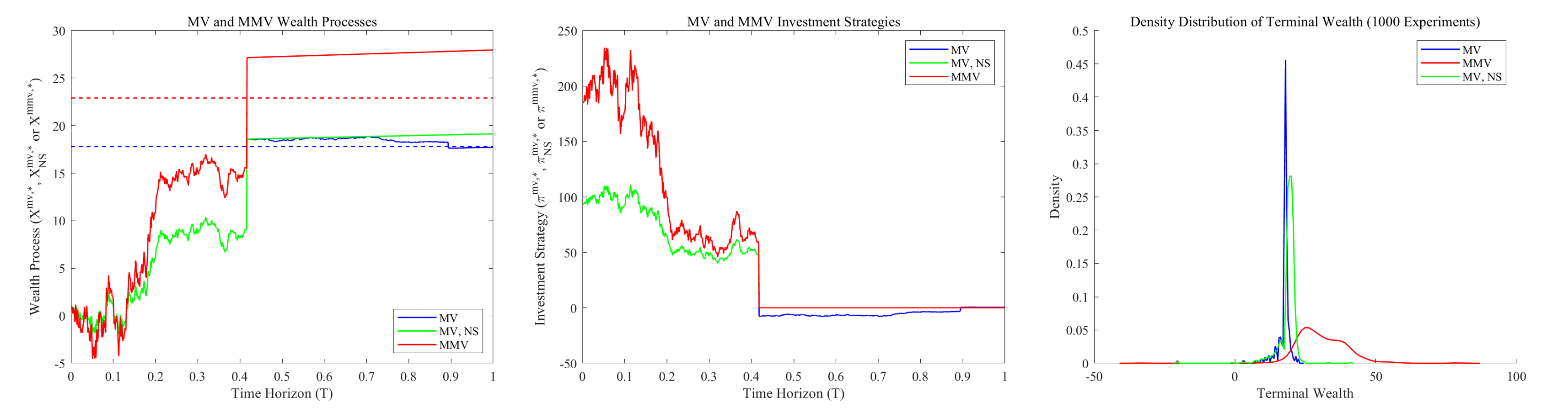}}
    {Constantly Distributed Jump \label{Constant Distributed}}
    {The first sub-figure shows the wealth process under the MV (MMV) optimal strategy, the second sub-figure shows the corresponding optimal strategy, and the third sub-figure shows the estimated density distribution (by 1000 experiments) of investor's terminal wealth. The solid blue curve represents the MV strategy, the green one signifies MV with no-short constraint, and the red curve corresponds to MMV. The blue (red) dashed curve represents the pre-determined target of wealth that drives the investor to short sell (hold zero position) of the risky asset under MV (MMV).} 
\end{figure}

\begin{figure}[htbp]
    \FIGURE
    {\includegraphics[width=1\linewidth]{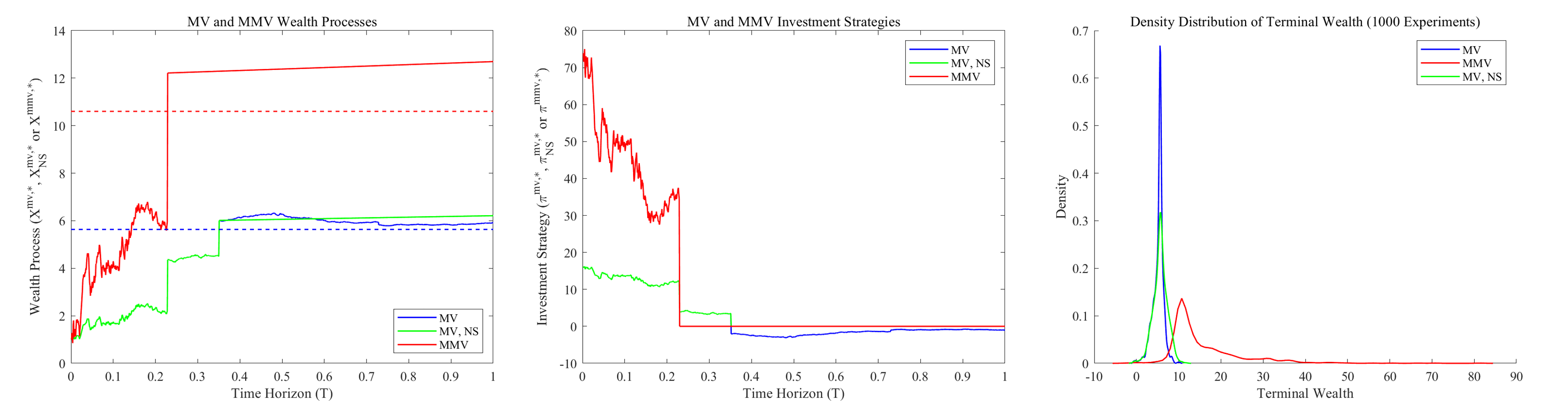}}
    {Uniformly Distributed Jump \label{Uniform Distributed}}
    {The first sub-figure shows the wealth process under the MV (MMV) optimal strategy, the second sub-figure shows the corresponding optimal strategy, and the third sub-figure shows the estimated density distribution (by 1000 experiments) of investor's terminal wealth. The solid blue curve represents the MV strategy, the green one signifies MV with no-short constraint, and the red curve corresponds to MMV. The blue (red) dashed curve represents the pre-determined target of wealth that drives the investor to short sell (hold zero position) of the risky asset under MV (MMV). }
\end{figure}

\begin{figure}[htbp]
    \FIGURE
    {\includegraphics[width=1\linewidth]{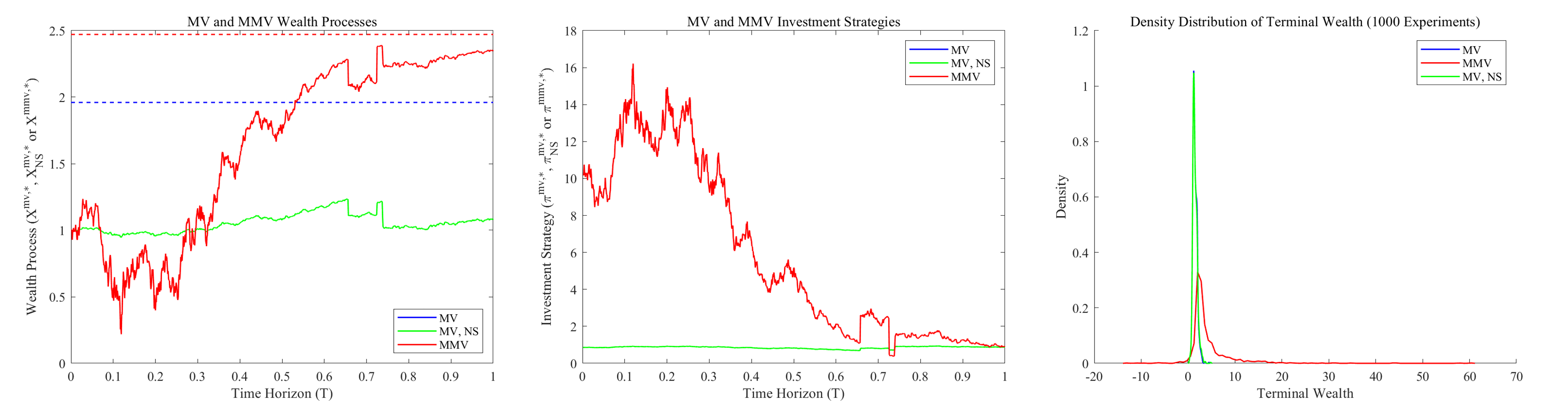}}
    {Exponential Distributed Jump \label{Exponential Distributed}}
    {The first sub-figure shows the wealth process under the MV (MMV) optimal strategy, the second sub-figure shows the corresponding optimal strategy, and the third sub-figure shows the estimated density distribution (by 1000 experiments) of investor's terminal wealth. The solid blue curve represents the MV strategy, the green one signifies MV with no-short constraint, and the red curve corresponds to MMV. The blue (red) dashed curve represents the pre-determined target of wealth that drives the investor to short sell (hold zero position) of the risky asset under MV (MMV).}
\end{figure}

Curves in Figure \ref{Exponential Distributed} simulate the dynamic wealth processes, the corresponding investment strategies and the wealth distribution. Little different from Figure \ref{Constant Distributed} or Figure \ref{Uniform Distributed}, jumps in this sample path are all smaller than $q^{mmv}$. Thus, investment strategies are all positive and wealth processes do not exceed the corresponding upper bound. The wealth process (investment strategy) under MV just completely coincides with that with no-short constraint. In this scenario, due to the low probability of the occurrence of relatively larger jumps, the wealth distribution under MV and that with no-short constraint are nearly identical. However, even if no significant jumps occur in actual sample paths, the expectation of such jumps cause the change in MMV investor's pricing operator, thereby resulting in the change in investor's end-of-period wealth distribution. 

\section{Conclusion}{\label{Section Conclusion}}
\noindent
In summary, we study the portfolio selection problem under the MMV preference proposed by \citet{maccheroni2009portfolio} in a jump-diffusion model and give an explicit solution different from that under the classical MV preference. Introducing the jump component necessitates the introduction of a new theorem, demonstrating that the target measures of MMV can be confined to the set of all non-negative Doléans-Dade exponentials. Then, we transform the original problem into a stochastic differential game and resolve it with the HJBI equation. Our investigation yields a pivotal sufficient and necessary condition: solutions to MV and MMV problems are consistent if and only if the financial market admits $\Delta{L} \leq q^{mv}$ a.s.. 

We reveal that the essential reason for the inconsistency of MV and MMV is that MMV investor takes a pricing operator which is always non-negative and thus avoids the problem of non-monotonicity. Such result is completely comparable to the earliest finding of \citet{dybvig1982mean}. 

Expanding our analysis to a multi-asset model, we validate the two-fund separation theorem and establish the monotone CAPM based on the market portfolio, which is a departure from the classical CAPM based on MV but comparable to the result in the single-period model derived by \citet{maccheroni2009portfolio}. Finally, we explore the MMV problem under a trading constraint and provide three specific numerical examples to illustrate our findings. We believe that empirical tests of MMV and monotone CAPM's effectiveness in real financial markets are worthy of future studies. 

\section*{Acknowledgements}
\noindent
The authors acknowledge the support from the National Natural Science Foundation of China (Grant Nos.12271290, 11871036). The authors also thank the members of the group of Mathematical Finance and Actuarial Sciences  at the Department of Mathematical Sciences, Tsinghua University for their feedback and useful conversations.

\newpage

\newpage
\section*{Figures}



\begin{APPENDICES}

\section{Proof of Theorem \ref{Theorem MMV Verification}} \label{Appendix Proof MMV Verification}
\noindent
For any $(\pi,\phi) \in \mathcal{A}^{mmv}_t \times \varPhi_t$ and arbitrarily large integer $N$, define a stopping time $T_N = T \wedge \inf\limits_{s \geq t}\big\{ \left|X^\pi(s)\right| \geq N \big\} \wedge \inf\limits_{s \geq t}\big\{ \left|Y^{\phi}(s)\right| \geq N \big\}$. Using  It\^{o}'s formula for the jump-diffusion process, we have
\begin{align}
    \mathbb{E}_{t,x,y}^{\mathbb{Q}^{\phi}} \Big[W \left( T_N,X^{\pi}\left(T_N\right) ,Y^{\phi}\left(T_N\right) \right)\Big] = W\left(t,x,y\right) +\mathbb{E}_{t,x,y}^{\mathbb{Q}^{\phi}} \Big[ \int_t^{T_N} \varPsi^{\pi,\phi}W \left(s,X^{\pi}\left(s\right),Y^{\phi}\left(s\right)\right)\mathrm{d}s \Big]. \notag
\end{align}
 
On the one hand, based on the assumption $\varPsi^{\pi^{\ast},\phi} W \left(s,X^{\pi^{\ast}}\left(s\right),Y^{\phi}\left(s\right)\right) \geq 0$, we obtain
\begin{equation}
    W\left(t,x,y\right) \leq \mathbb{E}_{t,x,y}^{\mathbb{Q}^{\phi}}\left[ W \left(T_N,X^{\pi^{\ast}}\left(T_N\right) ,Y^{\phi}\left(T_N\right) \right)\right]. \notag
\end{equation}
Letting $N \rightarrow \infty$ and using the second condition with the continuity of the function $W$, we have\\  $W(t,x,y) \leq J\left(t,x,y,\pi^{\ast},\phi\right) \leq \sup\limits_{\pi \in \mathcal{A}^{mmv}_t}J\left(t,x,y,\pi,\phi\right) $,\\
then
\begin{align}
    W(t,x,y) & \leq \inf_{\phi \in \varPhi_t}  \sup_{\pi \in \mathcal{A}^{mmv}_t}J\left(t,x,y,\pi,\phi\right), \notag \\
    W\left(t,x,y\right) & \leq  \inf_{\phi \in \varPhi_t}J\left(t,x,y,\pi^{\ast},\phi\right) \leq \sup_{\pi \in \mathcal{A}^{mmv}_t} \inf_{\phi \in \varPhi_t}J\left(t,x,y,\pi,\phi\right). \notag
\end{align}
On the other hand, as $\varPsi^{\pi,\phi^{\ast}} W \left(s,X^{\pi}\left(s\right),Y^{\phi^{\ast}}\left(s\right)\right) \leq 0$, we also have
\begin{align}
    W\left(t,x,y\right) \geq \mathbb{E}_{t,x,y}^{\mathbb{Q}^{\phi}}\left[ W \left( T_N,X^{\pi}\left(T_N\right) ,Y^{\phi^{\ast}}\left(T_N\right) \right)\right]. \notag
\end{align}
Similarly, we have $W(t,x,y) \geq J \left(t,x,y,\pi,\phi^{\ast} \right) \geq \inf\limits_{\phi \in \varPhi_t} J \left(t,x,y,\pi,\phi \right)$, then
\begin{align}
    & W(t,x,y) \geq \sup_{\pi \in \mathcal{A}^{mmv}_t}  \inf_{\phi \in \varPhi_t} J \left(t,x,y,\pi,\phi \right), \notag \\
    & W\left(t,x,y\right) \geq \sup_{\pi \in \mathcal{A}^{mmv}_t} J\left(t,x,y,\pi,\phi^{\ast} \right) \geq  \inf_{\phi \in \varPhi_t} \sup_{\pi \in \mathcal{A}^{mmv}_t} J\left(t,x,y,\pi,\phi \right). \notag
\end{align}
Combining the above inequalities proves the verification theorem. \hfill $\square$
\vskip 5pt
\section{Proof of Lemma \ref{Lemma Q Existence and Uniqueness}} \label{Appendix Proof Q Existence and Uniqueness}
\noindent
When $x\geq0$, 
\!\! \!\! \!\! \!\!\!\!\begin{align}
\!\!\!\!\!\!\!\!	\!\!\!\!\!\!{f}(x) & = \sigma^2 + \int_{-1}^{x} \widetilde{q}^2 \nu(\mathrm{d}\widetilde{q}) - \left(\mu-r + \int_{-1}^{x} \widetilde{q} \nu(\mathrm{d}\widetilde{q})\right) x \notag  =  \sigma^2 - (\mu-r)x - \lambda \mathbb{E}^\mathbb{P} \left[  (x - Q)Q \one_{\{-1 \leq Q \leq x\}} \right]. \notag
\end{align}
For any $x_2\geq x_1$, we have 
\begin{align}
	& \E^\mathbb{P}\left[(x_2 - Q)Q \one_{\{-1 \leq Q \leq x_2\}}\right]-\E^\mathbb{P}\left[(x_1 - Q)Q \one_{\{-1 \leq Q \leq x_1\}}\right] \notag \\
    = & \E^\mathbb{P}\left[(x_2 - Q)Q \one_{\{x_1 \leq Q \leq x_2\}}\right]+\E^\mathbb{P}\left[(x_2 - x_1)Q \one_{\{x_1 \leq Q \leq x_1\}}\right]. \notag
\end{align}
Then  
\begin{equation}
    \frac{\ud \mathbb{E}^\mathbb{P} \left[  (x - Q)Q \one_{\{-1 \leq Q \leq x\}} \right]}{\ud x}=\mathbb{E}^\mathbb{P} \left[ Q \one_{\{-1 \leq Q \leq x\}} \right]. \notag
\end{equation}
Thus, $f$ is continuous and its first order derivative is ${f}^{\prime}(x) = -(\mu-r) - \lambda \mathbb{E}^\mathbb{P} \left[ Q \one_{\{-1 \leq Q \leq x\}} \right].$
We  see that $f'$ is decreasing on $[0,\infty)$, which means that the function $f$ is either monotonically decreasing or increasing and then decreasing. Meanwhile, we have $f(0)=\sigma^2+\lambda\mathbb{E}^\mathbb{P} \left[ Q^2 \one_{\{-1 \leq Q \leq x\}} \right]>0,\ f(\infty)=-\infty$. This implies that $f$ admits a unique zero point $q^{mmv}$ on $[0,\infty)$.\\
\noindent	 
When $x\leq 0 $, 
\begin{align}
    {f}(x) & =  \sigma^2 - (\mu-r)x - \lambda \mathbb{E}^\mathbb{P} \left[  (x - Q)Q \one_{\{ Q > x\}} \right] \notag \\
	& = \sigma^2 - (\mu-r)x - \lambda \left(\mathbb{E}^\mathbb{P} \left[  (x - Q)Q\right]-\mathbb{E}^\mathbb{P} \left[  (x - Q)Q \one_{\{ -1\leq Q \leq x\}} \right]\right). \notag
\end{align}
Then $ f'(x)=-(\mu-r)-\lambda\mathbb{E}^\mathbb{P} \left[ Q \one_{\{ Q > x\}} \right]\leq -(\mu-r)-\lambda\mathbb{E}^\mathbb{P} \left[ Q \right]<0.$ Because $f(0)=\sigma^2+\lambda\mathbb{E}^\mathbb{P} \left[ Q^2 \one_{\{Q >x\}} \right]>0$, we have $f(x)>0$ on $[-1,0]$.
	 
In conclusion, we have that $f$ has a unique zero point $q^{mmv}$ on $[-1, \infty)$ and  $q^{mmv}\in(0, \infty)$. Further, simple calculation yields that Equation \eqref{Q Equation} holds. \hfill $\square$
\vskip 5pt
\section{Proof of Theorem \ref{Theorem MMV Function and Control}} \label{Appendix Proof MMV Function and Control}
\noindent
We only need to verify that the results given in Theorem \ref{Theorem MMV Function and Control} meet all the conditions in Theorem \ref{Theorem MMV Verification}. Directly substitute $\pi^*$ and $\phi^*$ into $\varPsi^{\pi,\phi}W\left(t,x,y\right)$, we have
\begin{equation}
    \inf_{\phi \in \varPhi_t} \varPsi^{\pi^{\ast},\phi}W\left(t,x,y\right) = 0, \quad \varPsi^{\pi^{\ast},\phi^{\ast}}W\left(t,x,y\right) = 0, \quad W\left(T,x,y\right) = x+y. \notag
\end{equation}
As we also have
\begin{equation}
    \sup_{\pi \in \mathcal{A}^{mmv}_t} \varPsi^{\pi,\phi^{\ast}}W\left(t,x,y\right) = \varPsi^{\pi^{\ast},\phi^{\ast}}W\left(t,x,y\right) = 0, \notag
\end{equation}
i.e.,  the first condition holds.

Under the control $\left(\pi^{\ast},\phi^{\ast}\right)$, the wealth process of the investor becomes
\begin{equation}
    \mathrm{d}X^{\pi^{\ast}}(s) =X^{\pi^{\ast}}(s^-)r \mathrm{d}s + 2e^{-(T-s)r}\left[e^{(T-s)C^{mmv}}Y^{\phi^{\ast}}(s^-) \mathrm{d} \widetilde{D}(s) - \mathrm{d}\left(e^{(T-s)C^{mmv}}Y^{\phi^{\ast}}(s^-)\right)\right], \notag
\end{equation}
where $\mathrm{d} \widetilde{D}(s) = \int_{q^{mmv+}}^{\infty} \left[ \frac{\left(\mu-r + \int_{-1}^{q^{mmv}} \widetilde{q} \nu(\mathrm{d}\widetilde{q})\right)q}{\sigma^2 + \int_{-1}^{q^{mmv}} \widetilde{q}^2 \nu(\mathrm{d}\widetilde{q})} - 1\right]\widetilde{N}(\mathrm{d}s,\mathrm{d}q)$. Define a stopping time $T_{q^{mmv}} = \inf \Big\{s > t: \int_{s^-}^s \int_{-1}^\infty \phi_2^{\ast}(q) N \left(\mathrm{d}\tau,\mathrm{d}q \right) \geq 1 \Big\} = \inf \Big\{s > t: \Delta{L}(s) \geq q^{mmv} \Big\}$,  $\inf \{\varnothing\}=\infty$. When $s < T_{q^{mmv}}$, there is no jump with size larger than $q^{mmv}$. Using the initial condition $X^{\pi^{\ast}}(t) = x,Y^{\phi^{\ast}}(t)=y$, we have 
\begin{equation}
     X^{\pi^{\ast}}(s) = e^{(s-t)r}x + 2e^{(T-t)C^{mmv}-(T-s)r}y - 2e^{(T-s)(C^{mmv}-r)}Y^{\phi^{\ast}}(s). \notag
\end{equation}
When $s = T_{q^{mmv}}$, we have 
\begin{align}
    \Delta X^{\pi^{\ast}}(T_{q^{mmv}}) = \pi^{\ast}(T_{q^{mmv}}) \Delta L(T_{q^{mmv}}) = 2e^{(T-T_{q^{mmv}})(C^{mmv}-r)} \frac{\Delta L(T_{q^{mmv}})}{q^{mmv}} Y^{\phi^{\ast}}(T_{q^{mmv}}^-), \notag
\end{align}
\begin{align}
    X^{\pi^{\ast}}(T_{q^{mmv}}) & = X^{\pi^{\ast}}(T_{q^{mmv}}^-)+\Delta X^{\pi^{\ast}}(T_{q^{mmv}}) \notag \\
    & =e^{(T_{q^{mmv}}-t)r}x + 2e^{(T-t)C^{mmv}-(T-T_{q^{mmv}})r}y \notag \\
    & + 2e^{(T-T_{q^{mmv}})(C^{mmv}-r)} \left(\frac{\Delta L(T_{q^{mmv}})}{q^{mmv}} - 1 \right) Y^{\phi^{\ast}}(T_{q^{mmv}}^-). \notag
\end{align}
When $T_{q^{mmv}} < s \leq T$, $Y^{\phi^{\ast}}(s^-) = 0$ and $\pi^{\ast}(s) = 0$, then we have
\begin{equation}
    X^{\pi^{\ast}}(s) = e^{(s-t)r}x + 2e^{(T-t)C^{mmv}-(T-s)r}y + 2e^{(T-T_{q^{mmv}})C^{mmv}- (T-s)r} \left(\frac{\Delta L(T_{q^{mmv}})}{q^{mmv}} - 1 \right) Y^{\phi^{\ast}}(T_{q^{mmv}}^-). \notag
\end{equation}
Now, we show that $\pi^*$ and $\phi^*$ is admissible. Because $\phi^*$ is bounded, we know that $Y^{\phi^*}$ is a square integrable martingale (see M\'ermin's criterion in \citet{memin2006decompositions}), and then $\phi^*$ is admissible. Moreover, using the useful moment estimate inequality for integrals with the L\'evy process (see, e.g., \citet{Protter2005}, Theorem 66 in Chapter 5), we have
\begin{equation}
    \E_{t,x,y}\Big[\sup_{t \leq s \leq u } \left| Y^{\phi^{\ast}}(s) \right|^8\Big]\leq K_1 \left(1+\int_{0}^{u}\E_{t,x,y}\Big[\sup_{t \leq s \leq r } \left| Y^{\phi^{\ast}}(s) \right|^8\Big]\ud r\right). \notag
\end{equation}
The Gronwall's inequality yields $\E_{t,x,y}\Big[\sup_{t \leq s \leq T } \left| Y^{\phi^{\ast}}(s) \right|^8\Big]<\infty.$ 
Based on the relationship between $X^{\pi^{\ast}}$ and $Y^{\phi^{\ast}}$, we have
\begin{equation}
    \begin{aligned}
    \mathbb{E}_{t,x,y}^{\mathbb{Q}^{\phi}} \Big[\sup_{t \leq s \leq T } \left| X^{\pi^{\ast}}(s) \right|^2 \Big] &\leq \sqrt{\mathbb{E}_{t,x,y}\Big[\left(\frac{\mathrm{d}\mathbb{Q^{\phi}}}{\mathrm{d}\mathbb{P}} \right)^2\Big] \mathbb{E}_{t,x,y} \Big[\sup_{t \leq s \leq T } {\left| X^{\pi^{\ast}}(s) \right|}^4 \Big]}  \\
    &\leq \sqrt{\mathbb{E}_{t,x,y}\Big[\left(\frac{\mathrm{d}\mathbb{Q^{\phi}}}{\mathrm{d}\mathbb{P}} \right)^2\Big] \left(K_2 + K_2 \mathbb{E}_{t,x,y} \Big[\sup_{t \leq s \leq T } {\left| Y^{\phi^{\ast}}(s) \right|}^8 \Big] + K_2 \E\left[ Q^8\right]\right)} < \infty. \notag
    \end{aligned}
\end{equation}
Thus, $\pi^{\ast} \in \mathcal{A}^{mmv}_t$ is an admissible trading strategy.

As for the second condition, for any $\phi \in \varPhi_t$, using the Cauchy's inequality and Doob's maximal inequality, we have
\begin{align}
    \mathbb{E}_{t,x,y}^{\mathbb{Q}^{\phi}} \Big[\sup_{t \leq s \leq T } \left| Y^{\phi^{}}(s) \right|\Big] \leq \sqrt{\mathbb{E}_{t,x,y}\Big[\left(\frac{\mathrm{d}\mathbb{Q^{\phi}}}{\mathrm{d}\mathbb{P}} \right)^2\Big] \mathbb{E}_{t,x,y} \Big[\sup_{t \leq s \leq T } {\left| Y^{\phi^{}}(s) \right|}^2 \Big]} < \infty. \notag
\end{align}
For any $\pi \in \mathcal{A}^{mmv}_t,\phi \in \varPhi_t$, using the Cauchy's inequality,
\begin{align}
    \mathbb{E}_{t,x,y}^{\mathbb{Q}^{\phi}} \Big[\sup_{t \leq s \leq T} \left|X^{\pi}(s)\right| \Big] \leq \sqrt{\mathbb{E}_{t,x,y}^{\mathbb{Q}^{\phi}} \Big[\sup_{t \leq s \leq T} {\left|X^{\pi}(s)\right|}^2 \Big]} < \infty. \notag
\end{align}
Then
\begin{equation}
    \mathbb{E}_{t,x,y}^{\mathbb{Q}^{\phi}} \left[ \sup_{t \leq s \leq T} \left| W \left(s, X^{\pi}\left(s\right) , Y^{\phi}\left(s\right) \right)  \right| \right] \leq e^{(T-t)r}\mathbb{E}_{t,x,y}^{\mathbb{Q}^{\phi}} \Big[\sup_{t \leq s \leq T} \left|X^{\pi}(s)\right| \Big] + e^{(T-t)C^{mmv}}\mathbb{E}_{t,x,y}^{\mathbb{Q}^{\phi}} \Big[\sup_{t \leq s \leq T} \left|Y^{\phi}(s)\right| \Big] < \infty. \notag
\end{equation}

To sum up, $W(t,x,y)$ and $(\pi^{\ast},\phi^{\ast})$ satisfy the conditions given by the verification theorem, and are respectively the value function and the optimal control of the robust control problem \eqref{MMV Auxiliary Function}. 

For the original MMV problem, by setting $y = \frac{1}{2\gamma}$, we hold the relationship
    \begin{align}
        X^{\ast}(s) & = e^{(s-t)r}x + \frac{1}{\gamma}e^{(T-t)C^{mmv}-(T-s)r} - 2e^{(T-s)(C^{mmv}-r)}Y^{\phi^{\ast}}(s) \one_{\{t \leq s < T_{q^{mmv}}\}} \notag \\
        & + 2e^{(T-T_{q^{mmv}})C^{mmv}- (T-s)r} \left(\frac{\Delta L(T_{q^{mmv}})}{q^{mmv}} - 1 \right) Y^{\phi^{\ast}}(T_{q^{mmv}}^-) \one_{\{s \geq T_{q^{mmv}}\}}. \notag 
    \end{align}
Then, the optimal investment strategy can be represented as 
    \begin{equation}
        \pi^{\ast}(s) = \frac{1}{q^{mmv}} \max\left\{ e^{(s-t)r}x + \frac{e^{(T-t)C^{mmv}-(T-s)r}}{\gamma} -X^{\ast}(s^-), 0\right\}, \notag
    \end{equation}
which proves Theorem \ref{Theorem MMV investor}. 
\hfill $\square$
\vskip 5pt

\section{Proof of Proposition \ref{Proposition Monotone Domain}} \label{Appendix Proof Monotone Domain}
\noindent
Based on Proposition \ref{Proposition MV Expression}, the terminal wealth under the MV optimal strategy satisfies
\begin{equation}
     X^{mv,\ast}(T) - \left(\mathbb{E}_{t,x}^\mathbb{P}\left[X^{mv,\ast}(T)\right] + \frac{1}{\gamma}\right) = - \frac{1}{\gamma} \mathcal{E} \Bigg( -\int_t^. \frac{\sigma}{q^{mv}} \mathrm{d}B(\tau) - \int_t^.\int_{-1}^{\infty} \frac{q}{q^{mv}} \widetilde{N}(\mathrm{d}\tau,\mathrm{d}q) \Bigg)_T, \notag
\end{equation}
which is non-positive a.s. when $\Delta{L} \leq q^{mv}$ a.s., and can be positive with a positive probability when $\Delta{L} \leq q^{mv}$ a.s. does not hold. For the MMV investor, based on Proposition \ref{Proposition MMV Expression}, the terminal wealth satisfies 
\begin{equation}
    X^{mmv,\ast}(T) - \left(\mathbb{E}_{t,x}^\mathbb{P}\left[X^{mmv,\ast}(T)\right] + \frac{1}{\gamma}\right) = -2Y^{\phi^{\ast}}(T) \leq 0, \notag 
\end{equation}
when $\Delta{L} \leq q^{mv}$ a.s.. However, when $\Delta{L} \leq q^{mv}$ a.s. does not hold, we have
\begin{align}
    & \mathbb{P}\left(X^{mmv,\ast}(T) - \left(\mathbb{E}_{t,x}^\mathbb{P}\left[X^{mmv,\ast}(T)\right] + \frac{1}{\gamma}\right) > 0 \right) \notag \\
    \geq & \mathbb{P}\left(X^{mmv,\ast}(T) - \left(\mathbb{E}_{t,x}^\mathbb{P}\left[X^{mmv,\ast}(T)\right] + \frac{1}{\gamma}\right) > 0 \ \cap \ T_{q^{mmv}} \leq T \right) \notag \\
    = & \mathbb{P}\Bigg(2e^{(T-T_{q^{mmv}})C^{mmv}} \left(\frac{\Delta L(T_{q^{mmv}})}{q^{mmv}} - 1 \right) Y^{\phi^{\ast}}(T_{q^{mmv}}^-) \one_{\{T \geq T_{q^{mmv}}\}} \notag \\
    & - 2 \mathbb{E}_{t,x}^\mathbb{P}\left[e^{(T-T_{q^{mmv}})C^{mmv}} \left(\frac{\Delta L(T_{q^{mmv}})}{q^{mmv}} - 1 \right) Y^{\phi^{\ast}}(T_{q^{mmv}}^-) \one_{\{T \geq T_{q^{mmv}}\}} \right] > 0 \Bigg) >0. \notag
\end{align}
Thus, the terminal wealth would fall outside the monotone domain with a positive probability. \hfill $\square$ 
\vskip 5pt

\section{Proof of Theorem \ref{Theorem MMV V3 Representation}} \label{Appendix Proof MMV V3 Representation}
\noindent
Following the procedures of Lemma \ref{Lemma V and V1}, Lemma \ref{Lemma V1 and V3} and Theorem \ref{Theorem MMV V2 Representation}, we only need to prove that for each $Y$ that $Y\geq \delta>0$, $\E\left[Y\right]=1$, there exists $\boldsymbol{\phi} \in \boldsymbol{\varPhi}$ such that $Y=\mathcal{E}(\boldsymbol{\phi})_T$. Using the representation theorem for multi-dimensional Brownian motions and Poisson random measures \citep{ito1956spectral, benth2003explicit}, for each $Y$ that $Y\geq \delta>0$, $\E\left[Y\right]=1$, there exists a predictable process $\boldsymbol{\phi}$ such that 
\begin{equation}
    Y=1-\sum_{j=1}^{d}\int_{0}^{T} \phi_{1j}(t)\mathrm{d}B_j(t) -\sum_{i=1}^{n} \int_{0}^{T}\int_{-\infty}^{\infty}\phi_{2i}(t,q)\widetilde{N}_i(\mathrm{d}t,\mathrm{d}q). \notag
\end{equation}
Let $Y(t)=1-\sum_{j=1}^{d}\int_{0}^{t} \phi_{1j}(s)\mathrm{d}B_j(s) -\sum_{i=1}^{n} \int_{0}^{t}\int_{-\infty}^{\infty}\phi_{2i}(s,q)\widetilde{N}_i(\mathrm{d}s,\mathrm{d}q)$. Then, 
\begin{equation}
    \Delta Y(t)=-\sum_{i=1}^{n}\phi_{2i}(t,\Delta L_i(t))\one_{\Delta L_i(t)\neq 0}. \notag
\end{equation}
Similar to Lemma \ref{Lemma V1 and V3}, we have
\begin{equation}
    \sum_{i=1}^{n}\phi_{2i}(t,\Delta L_i(t))\one_{\Delta L_i(t)\neq 0}\leq -\Delta Y(t)\leq Y(t^-). \notag
\end{equation}

For every $i\neq j$, we have $\left[L_i,L_j\right](t)=0$, as $L_i$ and $L_j$ are independent. However, as $L_i$ and $L_j$ are pure jump processes, we have $0 = \left[L_i,L_j\right](t)=\sum_{0\leq s\leq t}\Delta L_i(s)\Delta L_j(s)$, which indicates that almost surely, $\Delta L_i(t) \Delta L_j(t)=0$  for all $t\in[0,T]$. Therefore, there is no common jump time of $L_i$, $L_j$. Thus, for $j=1,\cdots,n$, we have $\phi_{2i}(t,\Delta L_i(t))\one_{\Delta L_i(t)\neq 0}\leq Y(t^-)$. Now, we let $\phi_{1j}^{\prime}(t)=\frac{\phi_{1j}(t)}{Y(t^-)}$ and $\phi_{2i}^{\prime}(t,q) = \min\left\{\frac{\phi_{2i}(t,q)}{Y(t^-)},1 \right\}$. Then we have
\begin{equation}
    \mathrm{d}Y^{\boldsymbol{\phi}}(t) = Y^{\boldsymbol{\phi}}(t^{-}) \left(-\sum_{j=1}^{d}\phi_{1j}^\prime(t)\mathrm{d}B_j(t) -\sum_{i=1}^{n} \int_{-\infty}^{\infty}\phi_{2i}^\prime(t,q)\widetilde{N}_i(\mathrm{d}t,\mathrm{d}q) \right), \quad Y^{\boldsymbol{\phi}}(0) = 1, \notag
\end{equation}
which satisfies the desired property.\hfill $\square$ 

\end{APPENDICES}

\end{document}